\documentclass[]{aiaa-tc}

\usepackage{subfigure}
\usepackage{subfigmat}
\usepackage[dvipsnames]{xcolor}
\usepackage{amsmath}
\usepackage{amssymb}
\usepackage{multicol}
 
 \usepackage{soul}
 \usepackage{rotating}
\usepackage{pdflscape}

\definecolor{blue}{rgb}{0, 0.4470, 0.7410}
\definecolor{red}{rgb}{0.8500, 0.1250, 0.0480}
\definecolor{orange}{rgb}{0.8500, 0.3250, 0.0980} 
\definecolor{yellow}{rgb}{0.9290, 0.6940, 0.1250}
\definecolor{purple}{rgb}{0.4940, 0.1840, 0.5560}
\definecolor{green}{rgb}{0.4660, 0.6740, 0.1880}
\definecolor{ltblue}{rgb}{0.3010, 0.7450, 0.9330}
\definecolor{dkred}{rgb}{0.6350, 0.0780, 0.1840}
\definecolor{gray}{rgb}{0.22, 0.22, 0.3}

\begin{document}

\title{Airfoil wake modification with Gurney flap at Low-Reynolds number}

\author{Muralikrishnan Gopalakrishnan Meena\thanks{Graduate Research Assistant, Department of Mechanical Engineering and Florida Center for Advanced Aero-Propulsion, Florida State University, mg15h@my.fsu.edu.}
\ and Kunihiko Taira\thanks{Associate Professor, Department of Mechanical Engineering and Florida Center for Advanced Aero-Propulsion, Florida State University, ktaira@fsu.edu.}\\
{\normalsize\itshape
Florida State University, Tallahassee, FL 32310, USA} \\
\and
Keisuke Asai\thanks{Professor, Department of Aerospace Engineering, Tohoku University.}\\
{\normalsize\itshape
Tohoku University, Sendai, Miyagi 980-8579, Japan}
}
\maketitle
\begin{abstract}
The complex wake modifications produced by a Gurney flap on symmetric NACA airfoils at low Reynolds number are investigated. Two-dimensional incompressible flows over NACA 0000 (flat plate), 0006, 0012 and 0018 airfoils at a Reynolds number of $Re = 1000$ are analyzed numerically to examine the flow modifications generated by the flaps for achieving lift enhancement.  While high lift can be attained by the Gurney flap on airfoils at high angles of attack, highly unsteady nature of the aerodynamic forces are also observed. Analysis of the wake structures along with the lift spectra reveals four characteristic wake modes (steady, 2S, P and 2P), influencing the aerodynamic performance. The effects of the flap over wide range of angles of attack and flap heights are considered to identify the occurrence of these wake modes, and are encapsulated in a wake classification diagram. Companion three-dimensional simulations are also performed to examine the influence of three-dimensionality on the wake regimes. The spanwise instabilities that appear for higher angles of attack are found to suppress the emergence of the 2P mode. The use of the wake classification diagram as a guidance for Gurney flap selection at different operating conditions to achieve the required aerodynamic performance is discussed.
\end{abstract}

\section*{Nomenclature}

\begin{multicols}{2}
\noindent\begin{tabular}{@{}lcl@{}}
\textit{$a$}  &=& Lift slope\\
\textit{$C_D$}  &=& Coefficient of drag \\
\textit{$C_L$}  &=& Coefficient of lift \\
\textit{$\overline{C_D}$}  &=& Time-averaged drag coefficient\\
\textit{$\overline{C_L}$}  &=& Time-averaged lift coefficient\\
\textit{$\overline{C_L/C_D}$}  &=& Time-averaged lift-to-drag ratio\\
\textit{$c$}  &=& Airfoil chord length \\
\textit{$F_x$}  &=& Drag on airfoil \\
\textit{$F_y$}  &=& Lift on airfoil \\
\textit{$f_s$}  &=& Vortex shedding frequency \\
\end{tabular}
\vfill
\columnbreak

\noindent\begin{tabular}{@{}lcl@{}}
\textit{$h$}  &=& Gurney flap height\\
\textit{$l^*$}  &=& Characteristic frontal length \\
\textit{$Re$}  &=& Chord based Reynolds number\\
\textit{$St$}  &=& Strouhal number \\
\textit{$u_\infty$}  &=& Free stream velocity\\
\textit{$\alpha$}  &=& Angle of attack \\
\textit{$\nu$}  &=& Kinematic viscosity\\
\textit{$\rho$}  &=& Fluid density\\
\textit{$\omega$}  &=& Vorticity\\
\end{tabular}
\end{multicols}

\section{Introduction}

The field of low-Reynolds number aerodynamics has received increased attention over the past couple of decades from the engineering and scientific communities for its fundamental importance as well as its applications in designing small-scale air vehicles \cite{Mueller01, Mueller:ARFM03, Pines:JA06}, understanding biological flights and swimming \cite{Dickinson:JEB93, Ellington:Nature96, Triantafyllou:ARFM00, Shyy:AIAAJ11}, and more recently uncovering Martian aerodynamics \cite{Fujita:AIAA12, Suwa:AIAA12, Nagai:AIAA13, Yonezawa:JSASS14, Munday:JA15}. For these applications, the basic understanding of lifting-body aerodynamics in Reynolds number range of $Re = 10^2$ to $10^5$ becomes important. Extensive research on this topic has been performed to examine the Reynolds number effect \cite{Mueller01, Garmann:PF11}, high angle-of-attack wake dynamics \cite{Taira:JFM09, PittFord:JFM13, Mancini:PF15, DeVoria:JFM17, Edstrand:JFM16}, wing performance \cite{Lissaman:ARFM83, Torres:AIAA05}, flow unsteadiness caused by wing maneuver \cite{Carr:JA88, Leishman06, Chen:PF10, Garmann:PF11, Ozen:JFM12, Jantzen:PF14, Jones:EF16}, and influence of external disturbances \cite{Leishman06}.  The effect of compressibility is also studied on low-Reynolds number flows, as Martian air vehicles will likely encounter a range of Mach numbers as they operate in the unique Martian atmosphere \cite{OwenJGR77, SeiffJGR77, Fujita:AIAA12, Suwa:AIAA12, Anyoji:AIAA14, Munday:JA15}.

Despite the prevalence of low-Reynolds number aerodynamics research on the aforementioned areas, there has been relatively little work performed on modifying the flow field around bodies to enhance the aerodynamic characteristics of lifting bodies or alter the behavior of their unsteady wakes in the context of flow control. While a number of work have focused on the use of flow control to suppress separation at moderate Reynolds numbers \cite{Seifert:AST04, Greenblatt:PAS00, Little:EF10, Whalen:AIAA15, Yeh:AIAA17}, smaller number of studies have investigated the use of flow control at very low Reynolds numbers to alter the dynamics of unsteady wakes \cite{Munday:PF13, Taira:AIAAJ09, Taira:AIAAJ10, Taira:AFC10}.  In this work, we consider the use of a simple passive flow control device, namely the Gurney flap \cite{Liebeck:JOA78}, on symmetric airfoils and examine its influence on modifying the wake dynamics and aerodynamic characteristics. We have selected the use of Gurney flap as a flow modification technique due to its simplicity and its demonstrated ability to modify the flow around wings in higher-Reynolds number applications.

Initial research on Gurney flaps were conducted by Liebeck \cite{Liebeck:JOA78}. The flaps were tested on a Newman airfoil with its aerodynamic forces analyzed over various angles of attack and flap heights. 
Liebeck hypothesized that the formation of two counter-rotating vortices at the downstream side of the Gurney flap to be responsible for the increase in aerodynamic performance of the airfoil. Later, confirmation of Liebeck's hypothesis and extension of the use of Gurney flaps on race cars were also performed \cite{Katz:JOFE89,Neuhart:NASA88}. Further research on the physics and vortex dynamics involved with the attachment of the Gurney flap to the trailing-edge of airfoils at high Reynolds number range of $Re = 10^4~\text{to}~10^6$ has been performed with emphasis on uncovering optimal flap height and deployment strategy at different flight conditions \cite{Jeffrey:JOA00,Li:FTC02,Oshiyama:JSASS15}. Various numerical and experimental studies have been performed in recent years to study the effects of virtual Gurney flaps \cite{Zhang:AIAAJ09,Feng:JFM15} which are deployed using dielectric-barrier-discharge plasma actuators with the added advantage of being an active flow control technique.

On the low-Reynolds number flow side, studies on wakes at very low Reynolds number of $Re = 1000$ have been performed in literature for airfoils without Gurney flap attached to the trailing-edge \cite{SMittal:CMAME94,Kurtulus:IJMAV15,Kurtulus:IJMAV16}. One of the few available literature on incompressible flow over airfoils with Gurney flap at $Re=1000$ was conducted numerically by Mateescu et al. \cite{Mateescu:AIAA14}. Symmetric and cambered NACA airfoils with Gurney flap attached to the trailing-edge were analyzed by the authors. A basic knowledge on the flow physics and aerodynamic forces was also reported. Applications of flow control with Gurney flaps deployed at the trailing edge at high $Re$ flows are examined in the aforementioned studies. On the contrary, the availability of such literature is limited for low $Re$ cases. Moreover, the behavior of far field wake structure for both low as well as high $Re$ flows are not well documented. 

In the present work, we perform an extensive parametric study to examine the influence of Gurney flap on the aerodynamic characteristics and two-dimensional wake patterns behind NACA 0000, 0006, 0012 and 0018 airfoils. In the section below, the computational approach and setup for the study is described. This is followed by the discussion of results from two-dimensional direct numerical simulations (DNS) in Section \ref{sec:results}, where we first classify the four characteristic wake modes observed. A detailed discussion of each wake mode is also provided. We then examine the effects of the Gurney flap on lift, drag and lift-to-drag ratio for all airfoils. We also provide in Section \ref{sec:3D} a brief analysis on the spanwise effects on the wake modes by performing companion three-dimensional simulations. Concluding remarks are offered at the end of the paper to summarize the findings and to discuss potential uses of Gurney flaps.

\section{Computational Approach}

\subsection{Two-dimensional DNS}

\begin{figure}[t]
\begin{center}
\includegraphics[width=1\textwidth]{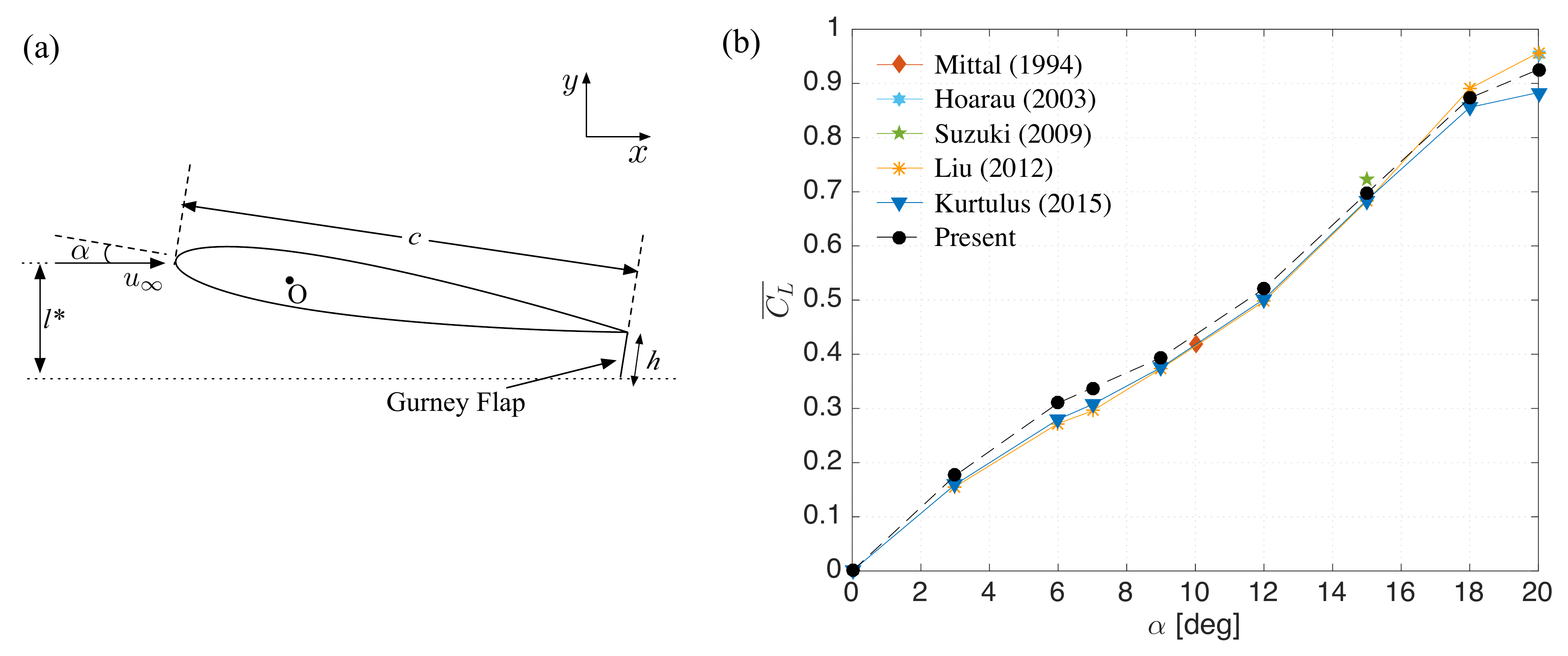} 
\end{center}
 \caption{(a) A representative setup with NACA 0012 airfoil at $\alpha=9^{\circ}$ with Gurney flap of $h/c = 0.1$ attached to the trailing-edge; (b) comparison of mean lift coefficient, $\overline{C_L}$, for NACA 0012 airfoil at $Re=1000$ with previous studies in literature \cite{SMittal:CMAME94,Hoarau:JFM03,Suzuki:EF09,Liu:CNSNS12,Kurtulus:IJMAV15}.}
 \label{f:setup}
\end{figure}

We investigate the influence of Gurney flaps on the two-dimensional wake behind four different symmetric airfoils of NACA 0000 (flat plate), 0006, 0012 and 0018 at a chord-based Reynolds number of $Re \equiv u_\infty c / \nu = 1000$.  Wide range of values for the Gurney flap height and angle of attack are considered in this study for the setup shown in Fig.~\ref{f:setup}(a). The flow field and force data obtained for each case are analyzed in detail to acquire a knowledge on the underlying effect of flow control from the use of the Gurney flap on the airfoils for aerodynamic performance enhancement.

The problem setup for the current work is depicted in Fig.~\ref{f:setup}(a).  Throughout this paper, the length, time, and velocity are non-dimensionalized by the chord length $c$, convective time scale $c/u_\infty$, and the free stream velocity $u_\infty$, respectively unless otherwise noted. Angles of attack $\alpha$ between $0^{\circ}$ to $20^{\circ}$ are considered for all the airfoils, of which Gurney flap is attached to the trailing-edge perpendicular to the chord line. For all cases, Gurney flap height of $h/c$ between $0$ to $0.15$ are considered. The wing is placed in the domain with its quarter chord at the origin with uniform flow prescribed at the inlet.

For the current analysis, the immersed boundary projection method \cite{Taira:JCP07,Colonius:CMAME08} is used to simulate the flow. This method is based on a finite volume formulation and incorporates the no-slip boundary condition along the immersed boundary into the projection operation. The scheme is second-order accurate in time and has a spatial accuracy of higher than first-order in the $L_2$ norm. Moreover, a multi-domain technique is used to simulate the flow over a body in free space. The scheme has been validated for a number of cases \cite{Taira:JFM09, Jantzen:PF14}, and has been found robust and accurate \cite{Colonius:CMAME08}. Five nested levels of multi-domains are used with the finest level being $(x/c,y/c) \in [-1,1] \times [-1,1]$ and the largest domain being $(x/c,y/c) \in [-16,16] \times [-16,16]$ in size. The time step for all cases is limited to a maximum CFL number of 0.3. 

In the current study, the drag and lift coefficients are defined as 
\begin{equation}
C_D \equiv \frac{F_x}{\frac{1}{2} \rho u^2_\infty c} \quad \text{and} \quad C_L \equiv \frac{F_y}{\frac{1}{2} \rho u^2_\infty c},
\end{equation}
respectively. The shedding frequency $f_s$ of lift is non-dimensionalized as the Strouhal number  
\begin{equation}
St \equiv \frac{f_s l^{*}}{u_\infty}, \label{eq:St}
\end{equation}
where the characteristic frontal length $l^*$ is taken to be
\begin{equation}
l^{*}= c\sin(\alpha) + h\cos(\alpha),
\end{equation}
as illustrated in Fig.~\ref{f:setup}(a). Grid convergence study was performed on the NACA 0012 airfoil without a flap at $\alpha = 10^\circ$ for grid size ranging from $200 \times 200$ to $500 \times 500$. A domain with $360 \times 360$ grid resolution was found to be sufficient with less than $1\%$ error in $|C_L|_{\max}$ and $St$ values, also capturing the wake structures effectively. Figure \ref{f:setup}(b) shows the comparison of mean $C_L$ values of NACA 0012 airfoil at $Re=1000$ with past studies \cite{SMittal:CMAME94,Hoarau:JFM03,Suzuki:EF09,Liu:CNSNS12,Kurtulus:IJMAV15}. The results obtained from the current simulations are in agreement with the data from literature.

\subsection{Three-dimensional DNS}
We also perform three-dimensional direct numerical simulations on selected cases from the two-dimensional analysis to acquire an understanding of the spanwise effects on the wakes. The analysis is performed on the NACA 0006 airfoil with Gurney flap of $h/c = 0.08$ and $\alpha = 6^\circ, 12^\circ$ and $18^\circ$ at $Re= 1000$. The cases are selected to represent the characteristic wake modes observed from the two-dimensional analysis. We use a finite-volume incompressible flow solver, {\tt Cliff} ({\tt CharLES} software package), developed by Stanford University and Cascade Technologies to solve the three-dimensional incompressible Navier-Stokes equation \cite{Ham:04,Ham:06,Morinishi:JCP98,Kim:JCP85}. A spatial domain of $(x/c,y/c,z/c) \in [-30,30] \times [-30,30] \times [0,4]$ is considered. The spanwise extent of $4c$ is chosen based on three-dimensional analysis for the baseline cases without Gurney flap at similar $Re$ range \cite{Neuhart:NASA88,Oshiyama:JSASS15}. A two-dimensional unstructured spatial grid discretization is utilized and extruded in the spanwise direction with $\Delta z/c = 0.0625$. The largest non-dimensional wall spacings along the suction side of the airfoil are $\Delta x = 0.004,~\Delta y = 0.007$ and $\Delta z = 0.0625$. The time step for all the simulations is set to a maximum CFL number of 1.0. No-slip boundary condition is specified on the airfoil and Gurney flap surfaces. An inflow boundary condition of $\mathbf{u}/u_\infty = (1,0,0)$ is prescribed at the inlet and symmetric boundary conditions at the far-field boundaries (top and bottom). A convective outflow boundary condition is specified at the outflow to allow the wake structures to leave the domain without disturbing the near-field solution. The flow is set to be periodic in the spanwise direction.

A three-dimensional perturbation is introduced to ensure the three-dimensional structures are sufficiently developed. Temporal convergence of the simulations is ensured through the rms and time-averaged values of pressure and velocities, obtained from a probe data located at the mid-span and $1c$ downstream of the trailing edge, having variations less than $2\%$ in time. The solver is validated with previous studies in literature \cite{Liu:CNSNS12,Kurtulus:IJMAV15} and current two-dimensional analysis for the NACA 0012 airfoil at $\alpha = 20^\circ$ with the time-averaged $\overline{C_L}$ having less than $2\%$ difference.

\section{Analysis of Two-Dimensional Flows} \label{sec:results}

Findings from the two-dimensional numerical experiments performed for flow over the four NACA airfoils with and without the Gurney flap attached to the trailing edge are discussed here. The addition of the Gurney flap and changes in angle of attack and airfoil thickness bring about various changes to the aerodynamic forces experienced by the airfoils. These variations in the aerodynamic forces are caused by the flow modifications to the airfoil wake. The observations on the airfoil wake modifications are first discussed followed by detailed analysis of the wake dynamics involved through the wake modifications. The effects on the aerodynamic forces are analyzed towards the end of this section.

\begin{figure}[t!]
\begin{center}
  	\begin{tabular}{|c|c|c|}\hline
      $\alpha $ &Instantaneous&Time-averaged \\ \hline \hline   
      	    $0^{\circ}$ &  \includegraphics[width=2.9in]{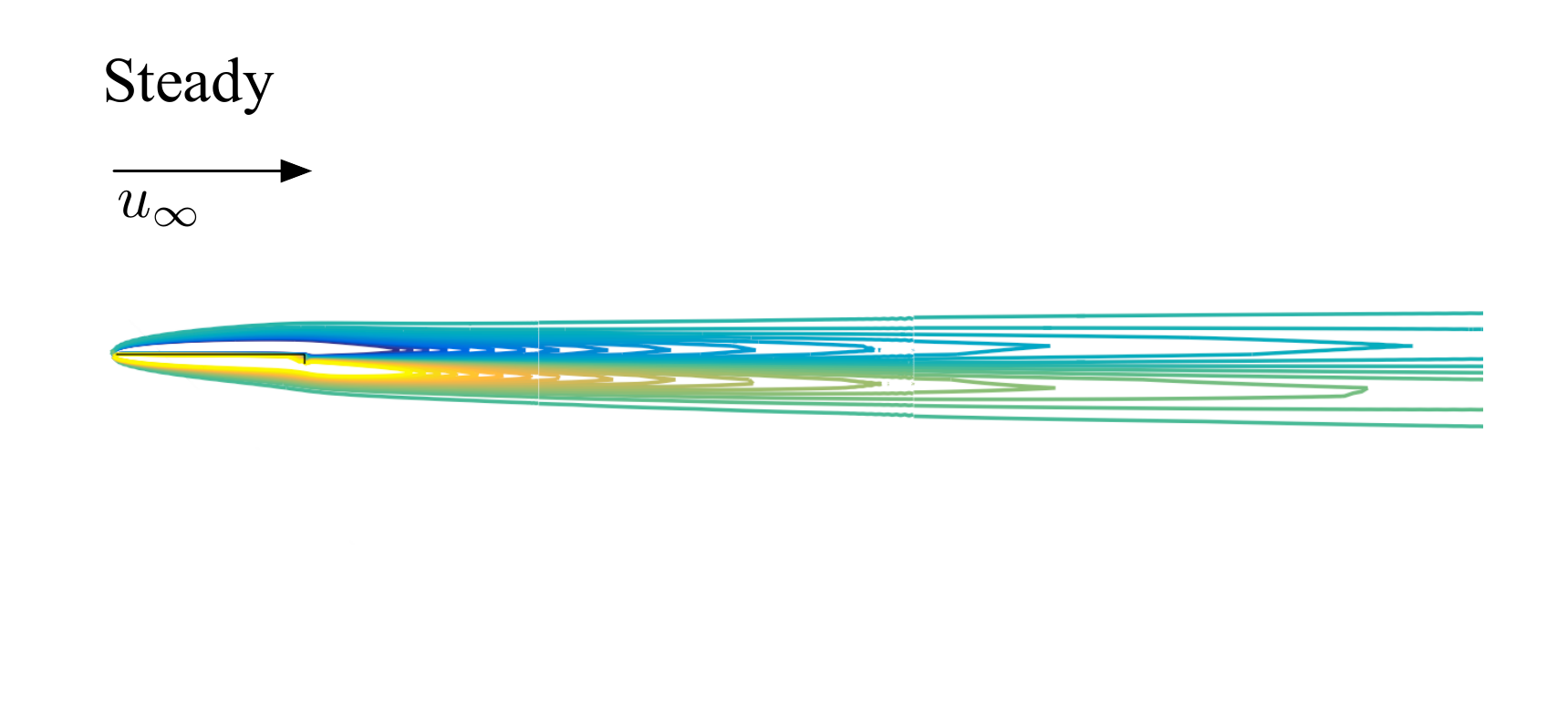}&  \includegraphics[width=2.9in]{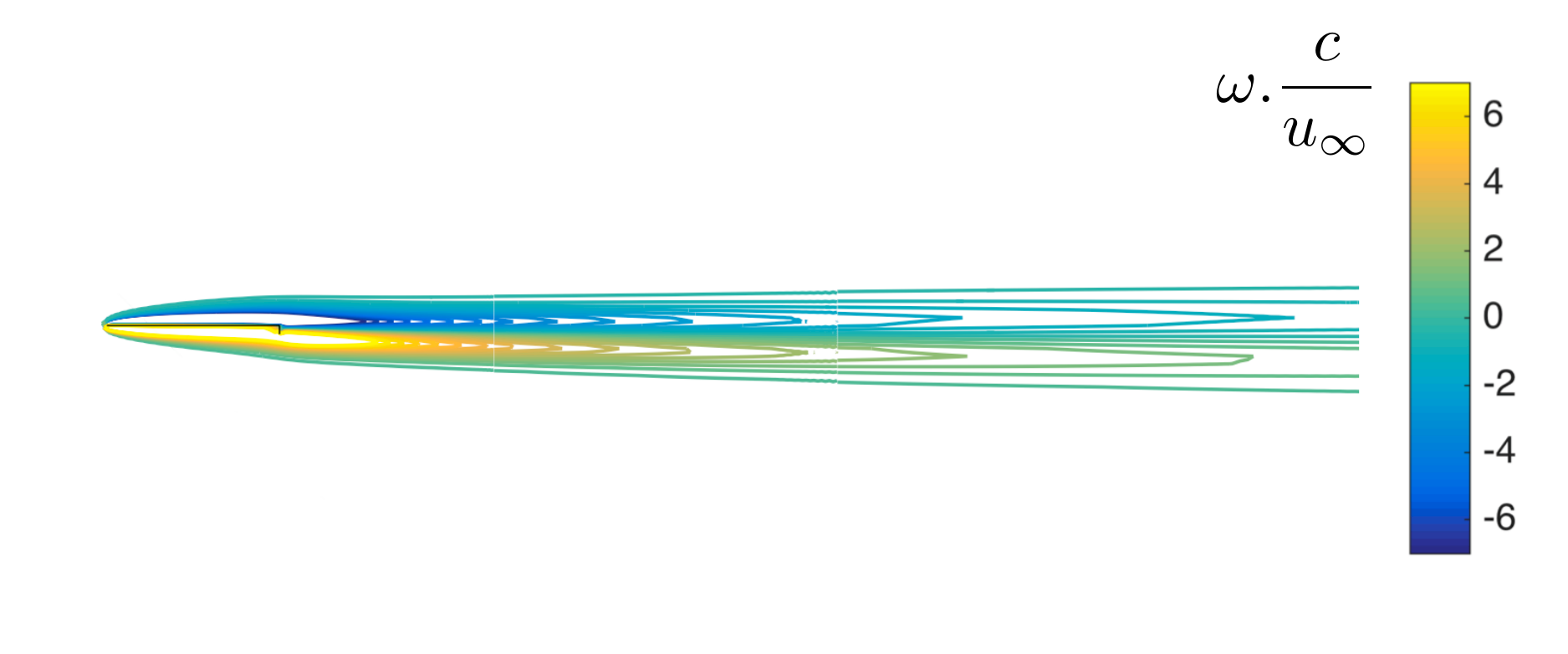}  \\ \hline        
            $6^{\circ}$ &  \includegraphics[width=2.9in]{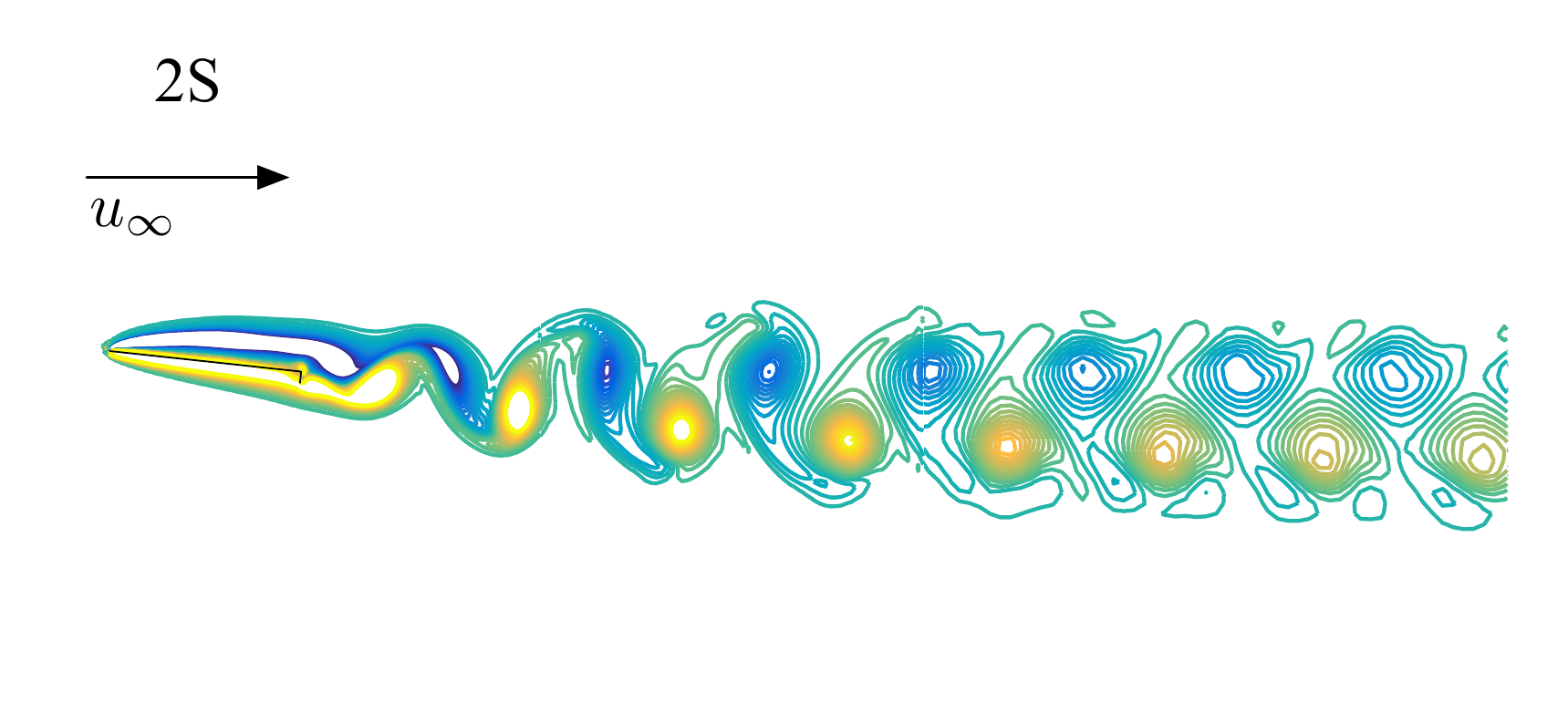}&  \includegraphics[width=2.9in]{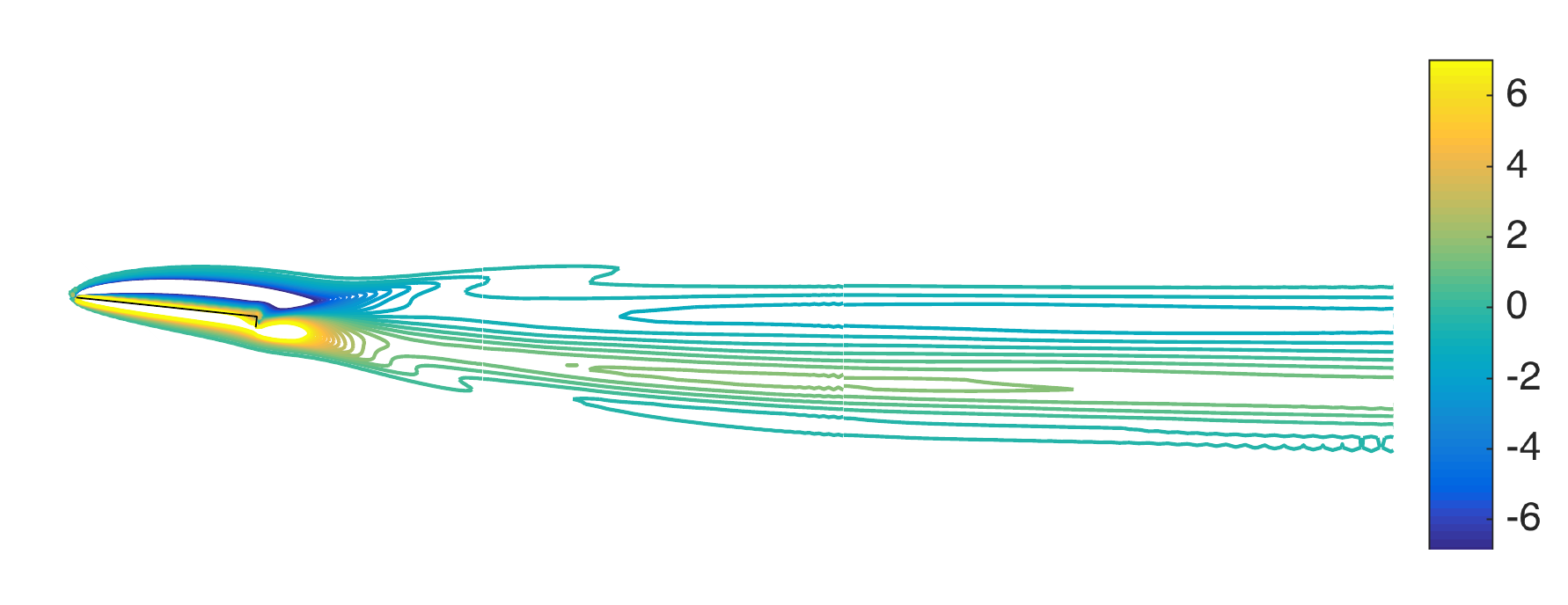}  \\ \hline
            $9^{\circ}$ &  \includegraphics[width=2.9in]{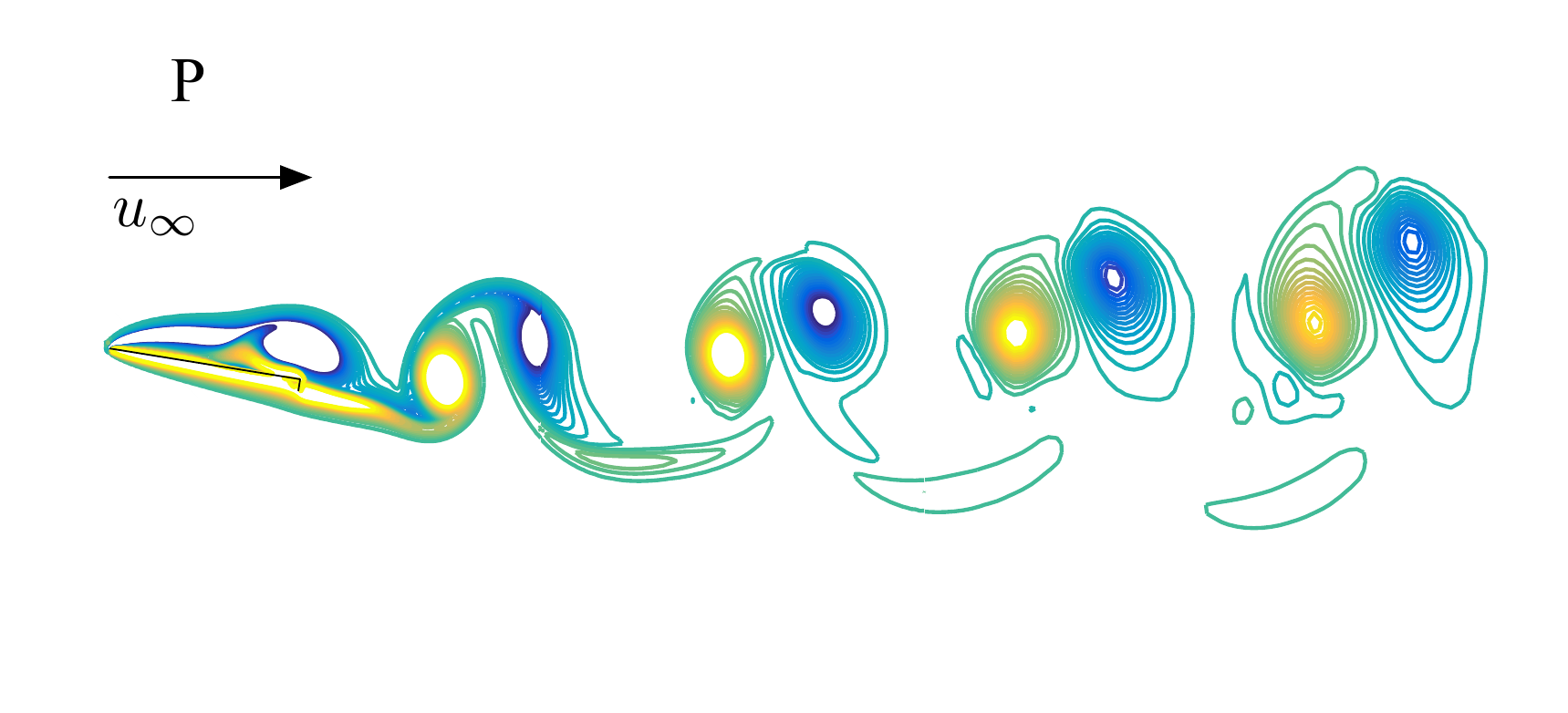}&  \includegraphics[width=2.9in]{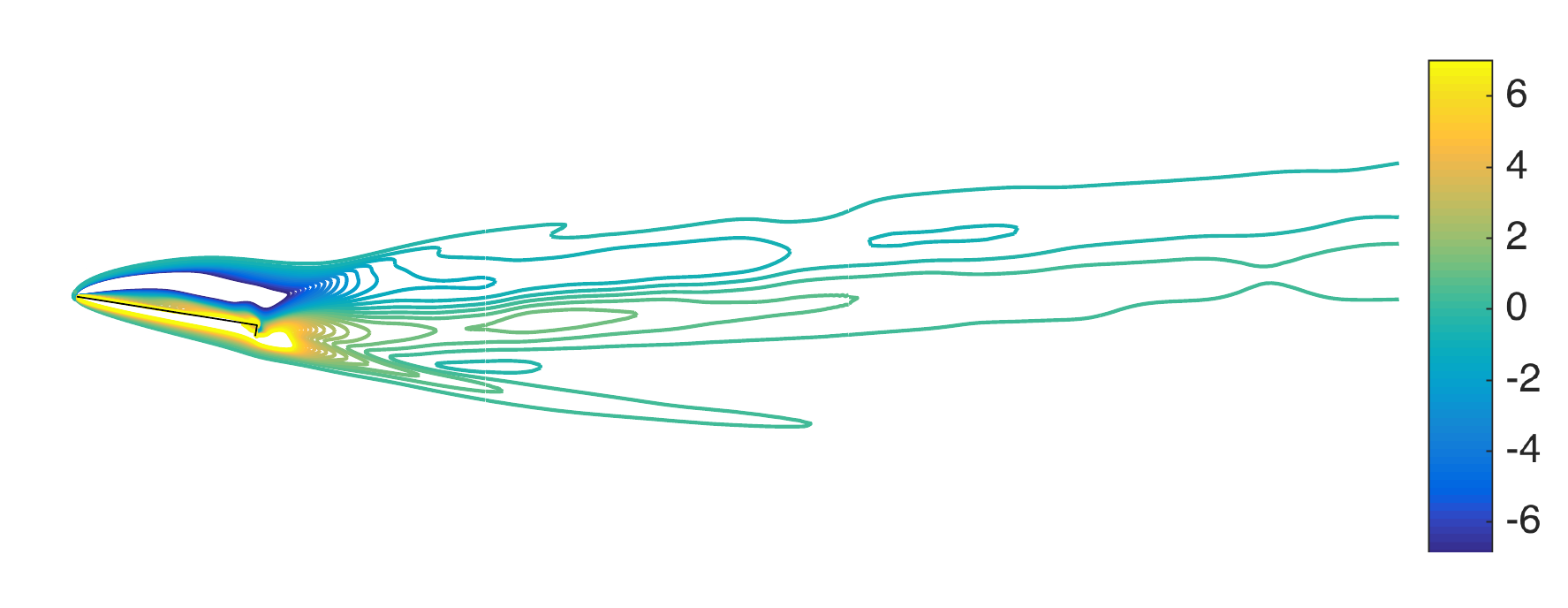}  \\ \hline
            $15^{\circ}$ &  \includegraphics[width=2.9in]{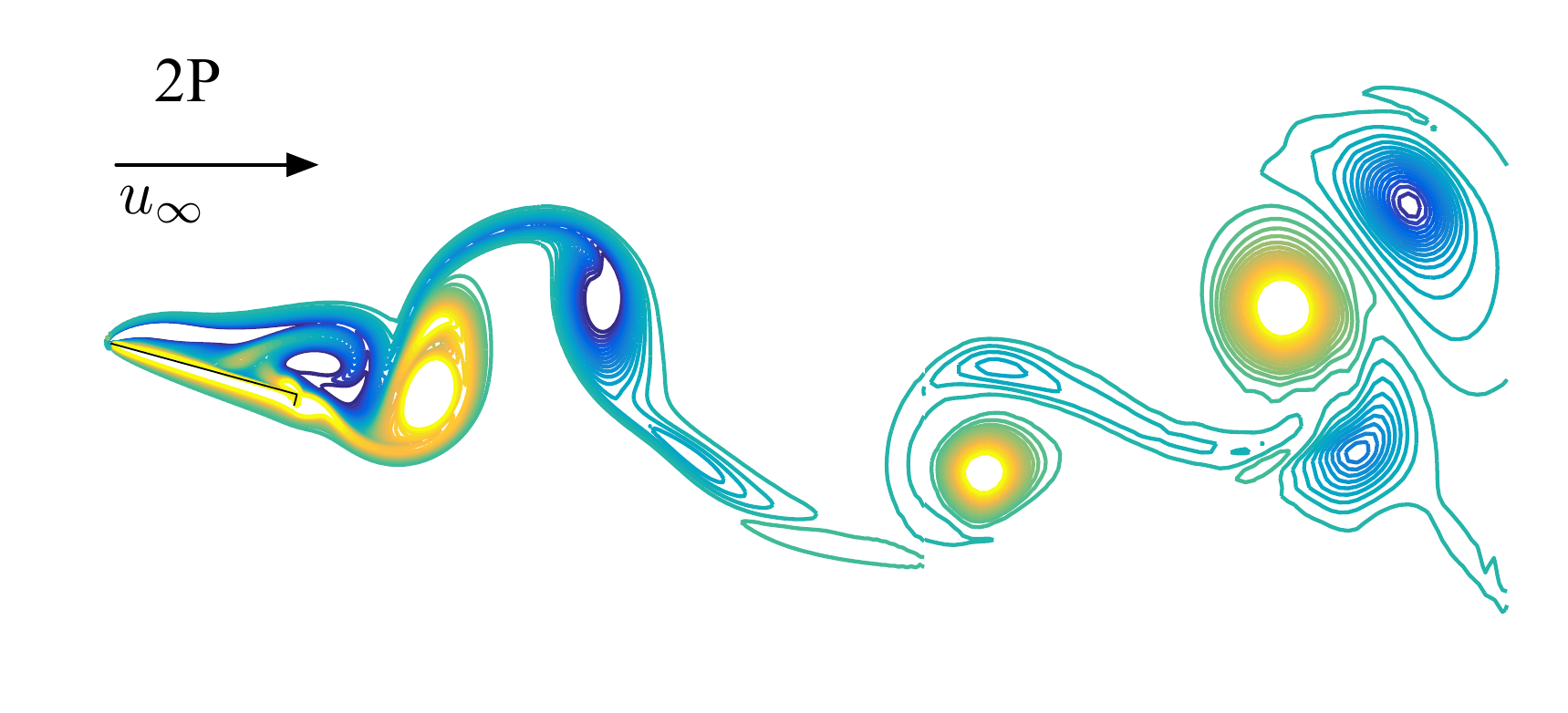}&  \includegraphics[width=2.9in]{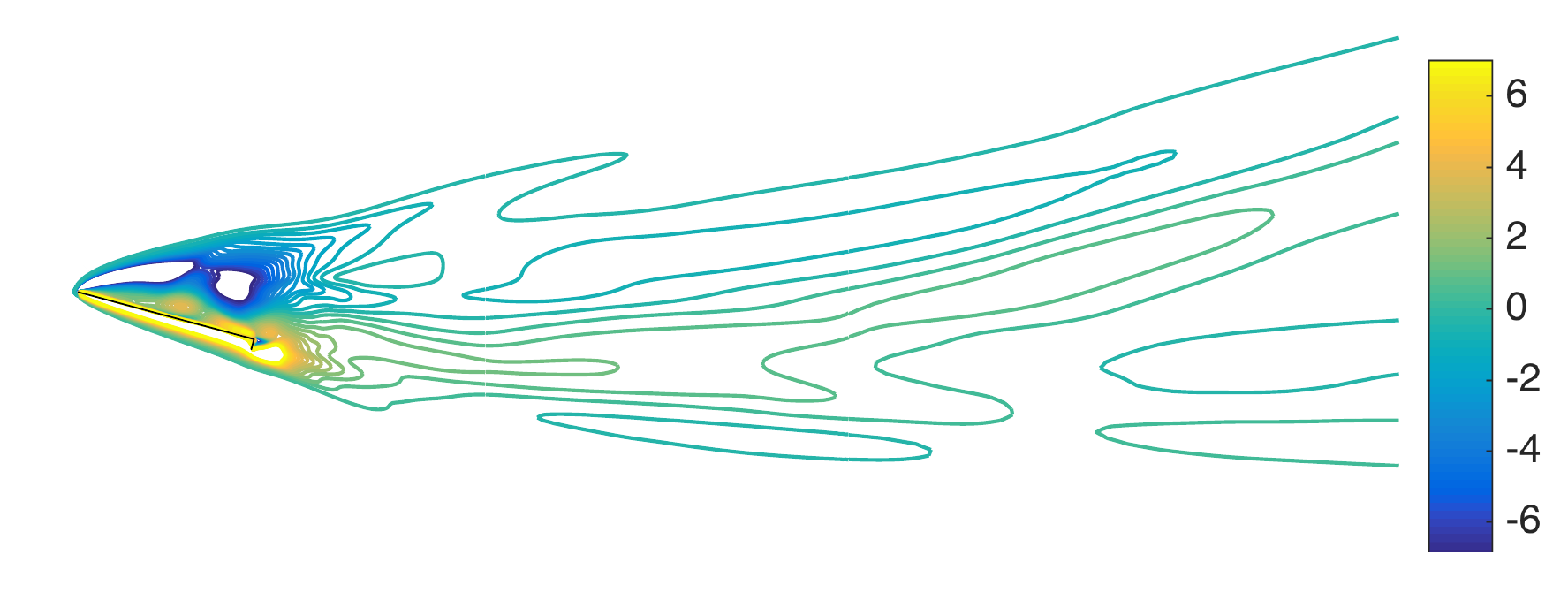}  \\ \hline
 
  \end{tabular}
\end{center}
\caption{Instantaneous and time-averaged two-dimensional vorticity fields $\omega c / u_\infty$, around NACA 0000 with a Gurney flap of $h/c = 0.06$ at various $\alpha$. The contour plots represent the four characteristic regimes: steady, 2S, P and 2P.}
\label{t:regime}
\end{figure}

\subsection{Wake mode classification}\label{sec:wake_classification}

First, let us analyze the vorticity field around the airfoils at $Re = 1000$ for various Gurney flap heights and angles of attack. We observe the formation of four distinct types of wakes develop over the airfoil both with and without the Gurney flap.

\begin{figure}[t!]
\begin{center}
\includegraphics[width=1\textwidth]{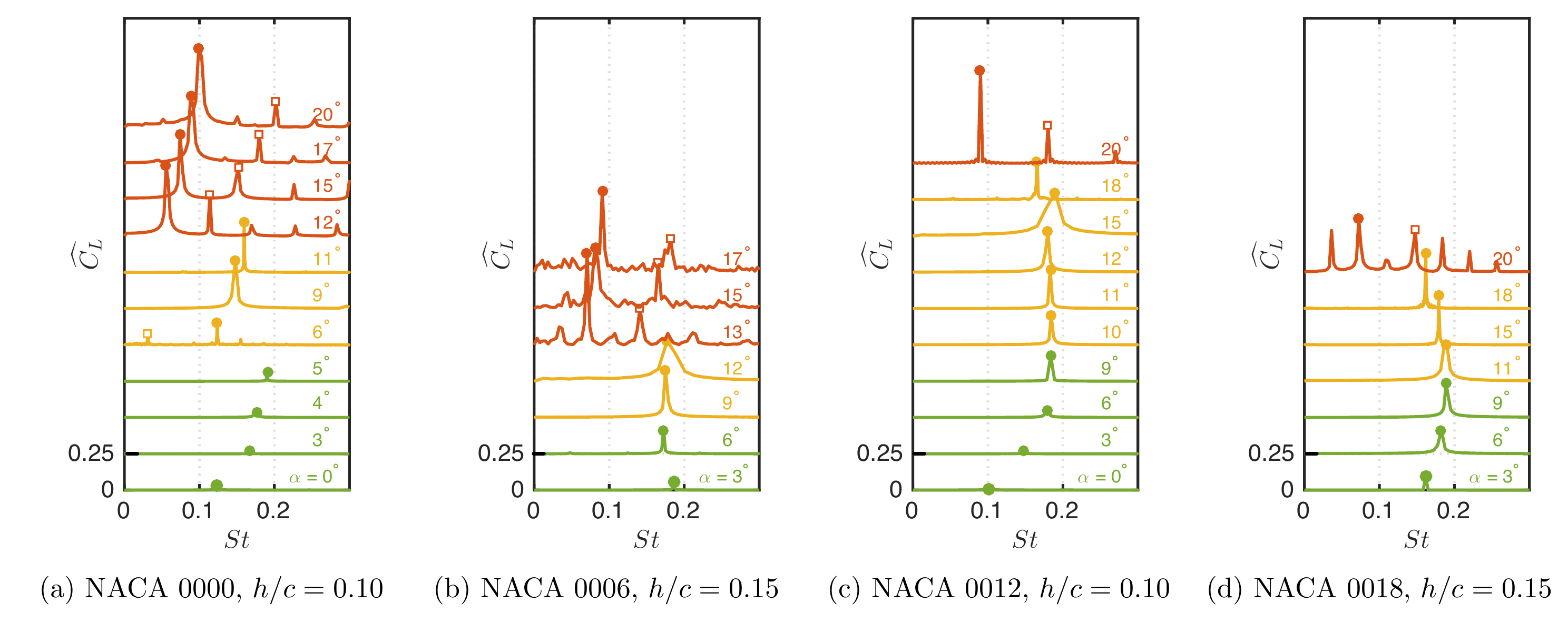} 
\end{center}
\caption{Frequency spectra for a range of $\alpha$ for NACA 0000, 0006, 0012, and 0018 airfoils with Gurney flap. Dominant vortex shedding frequency - {\large$\bullet$}, harmonic vortex shedding frequency - $\square$. Regions with dominant $St \in [0.12,0.18]$ are classified as 2S and P regimes by the \textcolor{OliveGreen}{green} and \textcolor{Dandelion}{yellow} regions respectively, and $St \in [0.06,0.10]$ as 2P regime by the \textcolor{Red}{red} region.}
 \label{f:spectra}
\end{figure}

The numerical simulations reveal that the flow fields can be characterized into four different regimes depending on the characteristics of vortex shedding.  The far field wake developments of different regimes for a representative airfoil setup (NACA 0000 with Gurney flap of $h/c = 0.06$) are summarized in Fig.~\ref{t:regime}. The four types of flow regimes observed are: 
\begin{itemize}
	\item Steady;
	\item 2S - periodic von K\'{a}rm\'{a}n vortex shedding;
	\item P - periodic shedding of single vortex pairs; and
	\item 2P - periodic shedding of two distinct vortex pairs.
\end{itemize}
These wake modes are classified on the basis of the flow structure of the wake, vortex shedding frequency and force fluctuations. In this section, we discuss the difference in the flow structure of the wake modes along with the vortex shedding frequencies corresponding to each mode. The frequency spectra of the lift history for selected cases of all airfoils are depicted in Fig.~\ref{f:spectra}. Sampling time sufficient to capture 30-50 dominant shedding cycles is used to perform the spectral analysis. The flow transition through the regimes are clearly evident by the abrupt changes in the $St$ values with $\alpha$ for NACA 0000 case.

The nomenclature of the modes follows the wake mode classification performed by Williamson and Roshko \cite{Williamson:JFS88} on the analysis of wakes behind a cylinder in forced oscillation. The wake modes observed in the current study correspond to the 2S, P and 2P wake modes defined by Williamson and Roshko. These 2S, P and 2P modes are previously observed in studies involving oscillating cylinder \cite{Ongoren:JFM88,Williamson:JFS88,Brika:JFM93} as well as airfoils \cite{Koochesfahani:AIAAJ89,Jones:AIAA96,Kurtulus:IJMAV15,Kurtulus:IJMAV16}. In particular, the wake modes have also been observed for the baseline case of NACA 0012 airfoil at $Re = 1000$ by Kurtulus \cite{Kurtulus:IJMAV15,Kurtulus:IJMAV16}.

In the steady regime (Fig.~\ref{t:regime}, $\alpha = 0^{\circ}$), the wake is steady and the flow is attached to the airfoil surface. This flow regime attributes to the lowest magnitude of drag experienced by the airfoil compared to that in the other unsteady modes. With increase in $\alpha$ or $h/c$ values, the flow starts to become unsteady. With this, the wake is classified into the next regime, the 2S mode. 

The 2S mode (Fig.~\ref{t:regime}, $\alpha = 6^{\circ}$) is characterized by unsteady flow with periodic von K\'{a}rm\'{a}n vortex shedding of alternating clockwise and counter-clockwise rotating vortices (2S represents two single vortices). The time-averaged vorticity contour lines are below the center of the wake. The wake height, observable from the time-averaged vorticity fields, is larger compared to that of the steady regime denoting increase in drag experienced by the airfoil. The smaller wake height for the 2S regime compared to the other unsteady regimes is also noticed. Also, a single peak in the frequency spectra, corresponding to the vortex shedding phenomena, for the cases classified into the 2S regime can be observed in Fig.~\ref{f:spectra}. The $St$ values corresponding to the dominant vortex shedding frequency lie between $[0.12,0.18]$ with increase in angle of attack through this regime. With further increase in $\alpha$ and $h/c$ values, the wake transitions to the next regime, the P mode.

The P mode (Fig.~\ref{t:regime}, $\alpha = 9^{\circ}$) is distinguished by periodic von K\'{a}rm\'{a}n shedding of a single vortex pair (P represents a single vortex pair). The spatial separation between two vortex pairs is distinctly larger compared to that observed between the single vortices in the 2S mode. The time-averaged vorticity contour lines of the regime shifts above the center of the wake, portraying decrease in lift experienced by the airfoil. Height of the wake increases considerably compared to the 2S regime. This increases the drag force experienced by the airfoil. Similar to the 2S regime, the cases classified in the P regime are characterized by the occurrence of single peaks in the frequency spectra as seen from Fig.~\ref{f:spectra}. The $St$ corresponding to these dominant vortex shedding frequency saturates to values in $[0.12,0.18]$ for this regime. Further increase in $\alpha$ and $h/c$ values causes the wake transition to the complex 2P mode.

The 2P mode (Fig.~\ref{t:regime}, $\alpha = 15^{\circ}$) is characterized by two pairs of vortices convecting above and below the center of the wake (2P represents two vortex pairs). The time-averaged vorticity field for the 2P regime evidently depicts the vortex pairs convecting away from the center of the wake. Compared to all other regimes, the 2P wake is most prominent and has the largest height amongst the four wake regimes. Consequently, the drag force experienced by the airfoils for this regime is the highest compared to that of all the other modes. Moreover, the downward force acting on the airfoil due to the upward displacement of the mean flow reduces the lift enhancement. The frequency spectra for cases classified in the 2P regime exhibits two prominent peaks, portraying the shedding of the two pairs of vortices. Also, the $St$ values corresponding to the dominant vortex shedding frequency drop sharply from $[0.12,0.18]$ and saturates at values $St \in [0.06,0.1]$. These observations from the frequency spectra of lift compared to that of the 2S and P modes signify the occurrence of the 2P mode. The occurrence of the 2P mode is found in baseline cases (without Gurney flap) for NACA 0012 at $Re=1000$ by Kurtulus \cite{Kurtulus:IJMAV15,Kurtulus:IJMAV16}, but only at very high angles of attack, $\alpha \in [23^{\circ},41^{\circ}]$. By adding the Gurney flap, emergence of the 2P regime is seen for lower $\alpha$. This 2P mode may be a predominantly two-dimensional phenomenon and may be suppressed by spanwise (three-dimensional) instabilities. We provide discussions on the influence of three-dimensionality on the 2P wake in Section \ref{sec:3D}. Thus, the emergence of the 2P mode should be used as a cautionary guide for the need of three-dimensional simulations to further analyze the flow. In fact, one should avoid operating an airfoil in such 2P regime in most cases, due to the large-scale unsteadiness. Nonetheless, we provide discussions on the wake dynamics and aerodynamic characteristics of the 2P mode in Sections \ref{sec:near_wake} and \ref{sec:aero_forces} to paint a complete description of the four wake regimes observed in two-dimensional flows.

\begin{figure}[t!]
 \begin{subfigmatrix}{2}
  \subfigure[NACA 0000]{\includegraphics{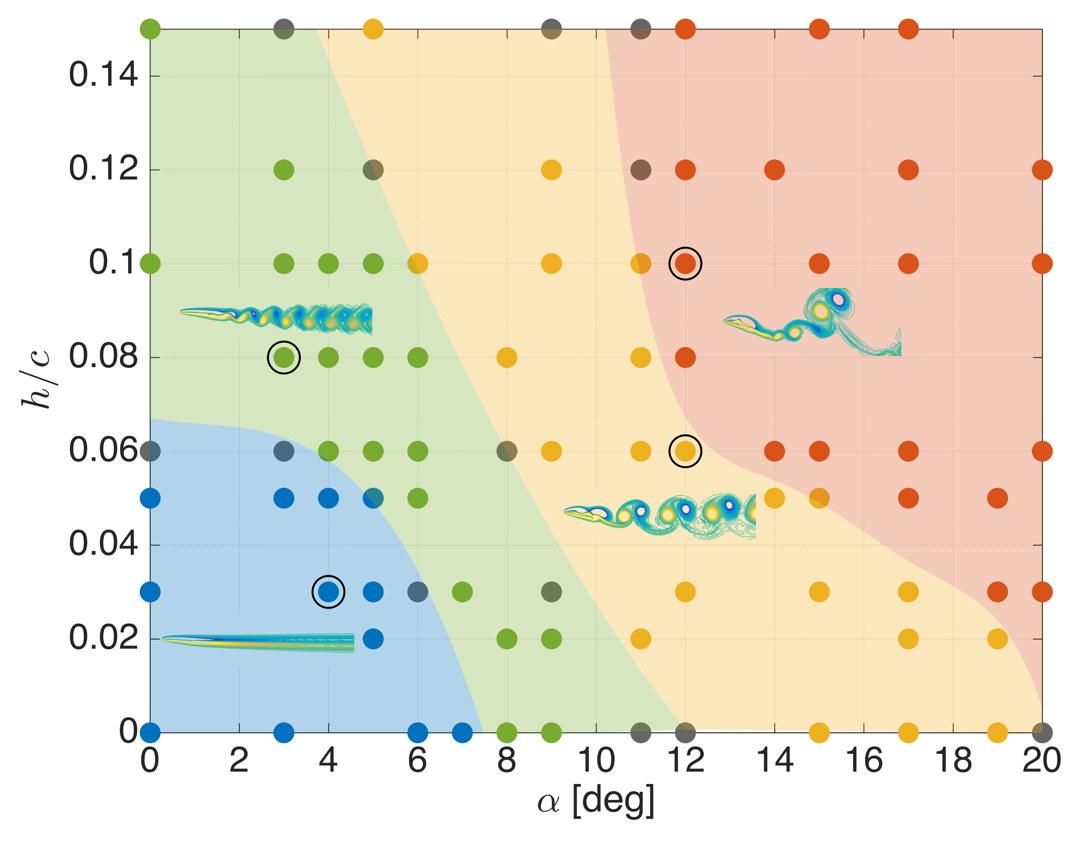}}
  \subfigure[NACA 0006]{\includegraphics{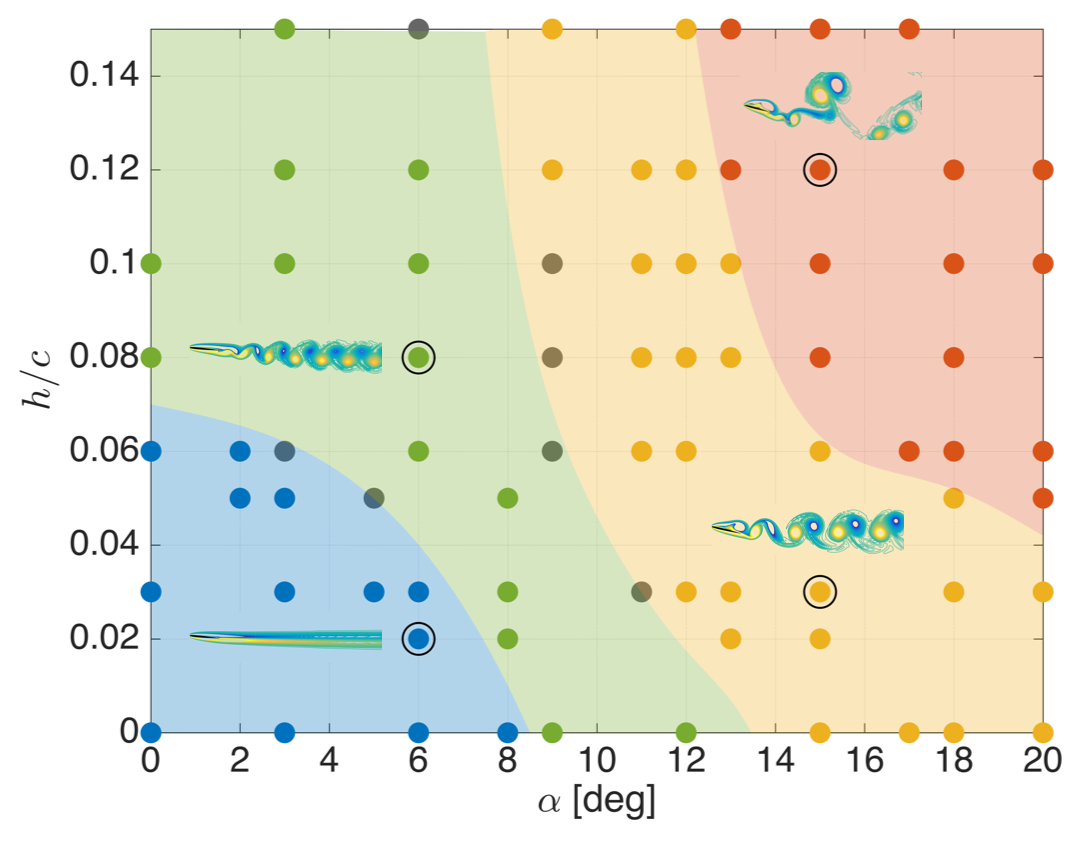}}
  \subfigure[NACA 0012]{\includegraphics{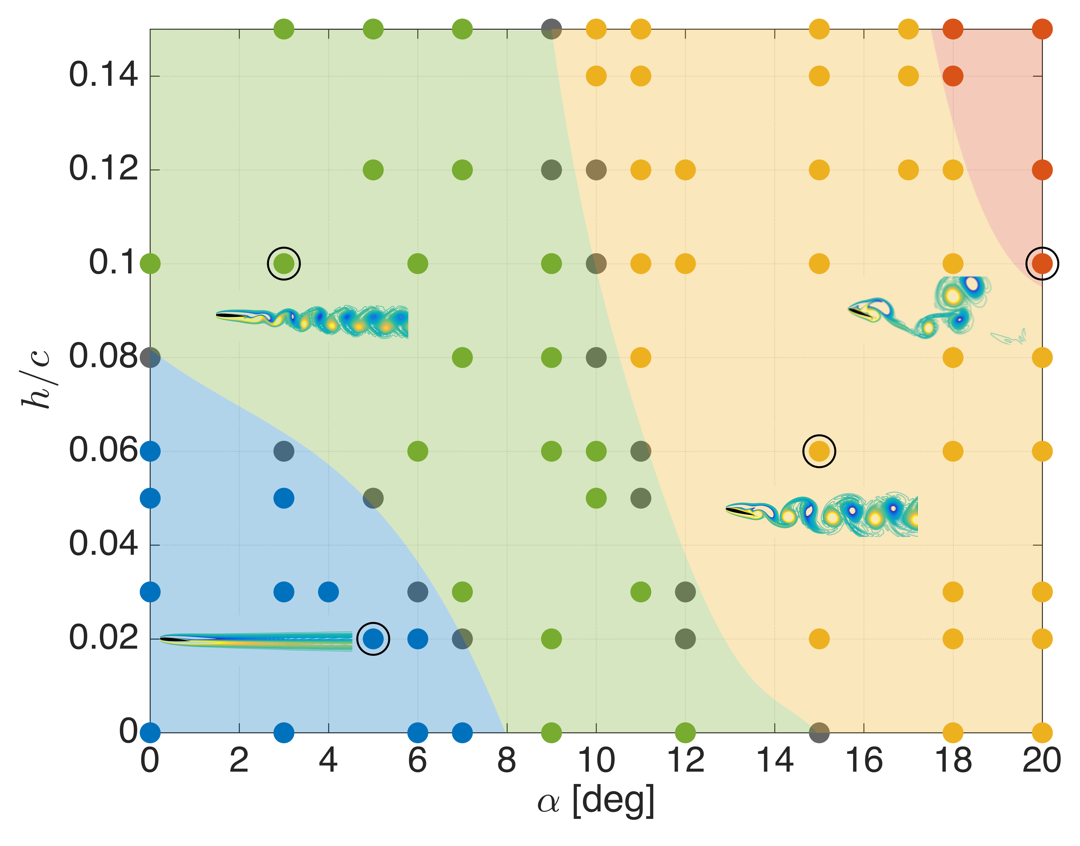}}
  \subfigure[NACA 0018]{\includegraphics{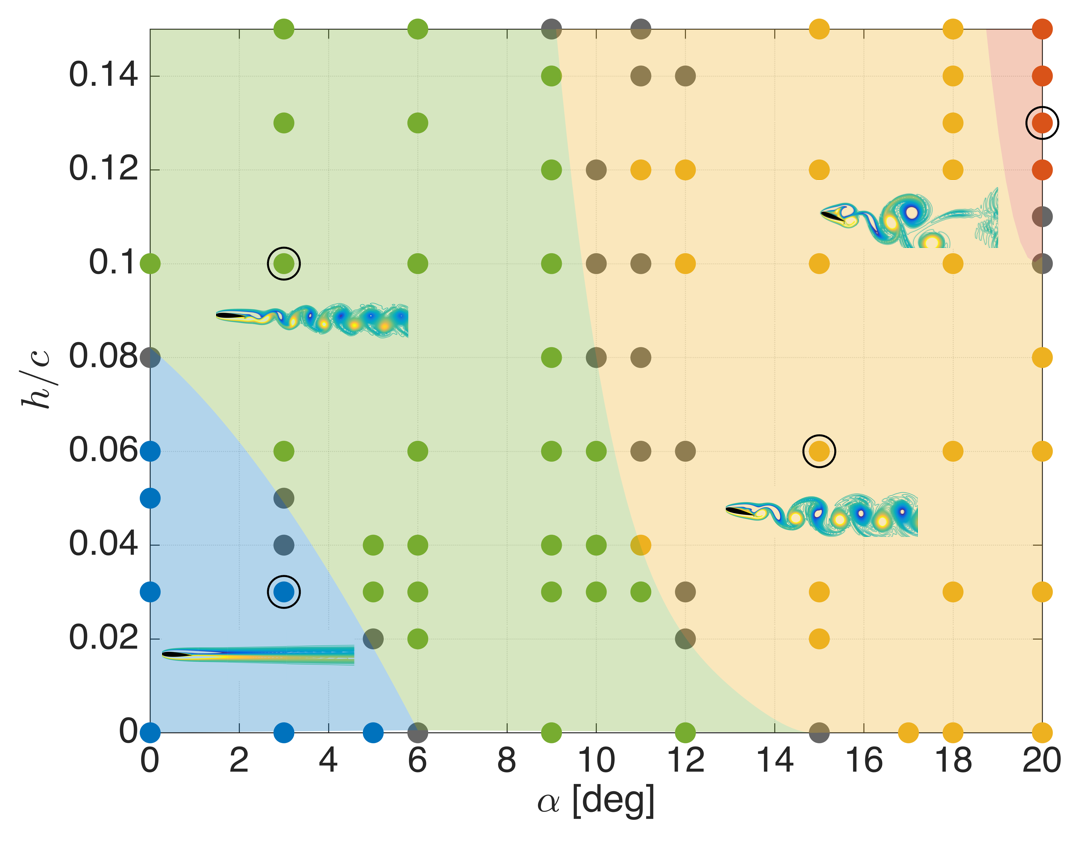}}
 \end{subfigmatrix}
  \caption{Wake classification diagram for NACA 0000, 0006, 0012, and 0018 airfoils are illustrated. All cases simulated in the current study are categorized into different characteristic wake regimes respect to $h/c$ and $\alpha$ values. Different regimes are described by: steady - \textcolor{NavyBlue}{blue}, 2S - \textcolor{OliveGreen}{green}, P - \textcolor{Dandelion}{yellow}, 2P - \textcolor{Red}{red} and transition between two regimes - \textcolor{Gray}{gray}. The boundaries for all the regimes are obtained by polynomial curve fitting.}
 \label{f:stability}
\end{figure}

We use the examination of the flow structure of the wake and the lift spectra to summarize the wake regimes of all four airfoils using a wake classification diagram on the $(\alpha,h/c)$ plane. Figure \ref{f:stability} provides an overview of the simulations performed and the influence of the Gurney flap on the wake modes for different airfoils. The baseline cases without flap are plotted on the $x$-axis. Representative vorticity fields of different wake regimes observed are also portrayed for each airfoil case. With the addition of the Gurney flap, transition of the unsteady wake modes is shifted to lower angles of attack. The protruding flap disrupts the inflow, causing the flow to be unsteady at lower angles of attack. The wake classification diagram shifts towards the left axis of the ($\alpha,h/c$) plane with increase in the Gurney flap height. 

When the airfoil geometry is changed from flat plate to a thick airfoil, the occurrence of the 2P regime is delayed to higher $h/c$ and $\alpha$ values. Furthermore, the size of the steady regime decreases. Increasing the thickness of the airfoil leads to a decrease in the influence of the Gurney flap as the thickness of the airfoil overshadows the effect of the flap. As a result of this, the complex 2P mode is revealed only at higher $\alpha$ and $h/c$ values for thicker airfoils.

The wake classification diagram provides a broad picture of the effect of adding Gurney flap to the trailing edge of the airfoil. The diagram is useful to determine the nature of flow at different $h/c$ and $\alpha$ values for all the airfoils studied. The overall performance of different airfoil-Gurney flap configurations can be understood from the diagram. This can be used to determine the optimal geometric and angle of attack conditions for different aerodynamic performance requirements. Depending on the necessity, the ideal airfoil and corresponding Gurney flap configuration can be chosen. In Section \ref{sec:3D}, we present insights from companion three-dimensional simulations and the influence of three-dimensionality on the wake dynamics.

\subsection{Near-wake Dynamics}\label{sec:near_wake}

\begin{figure}[t!]
\begin{center}
\includegraphics[width=0.9\textwidth]{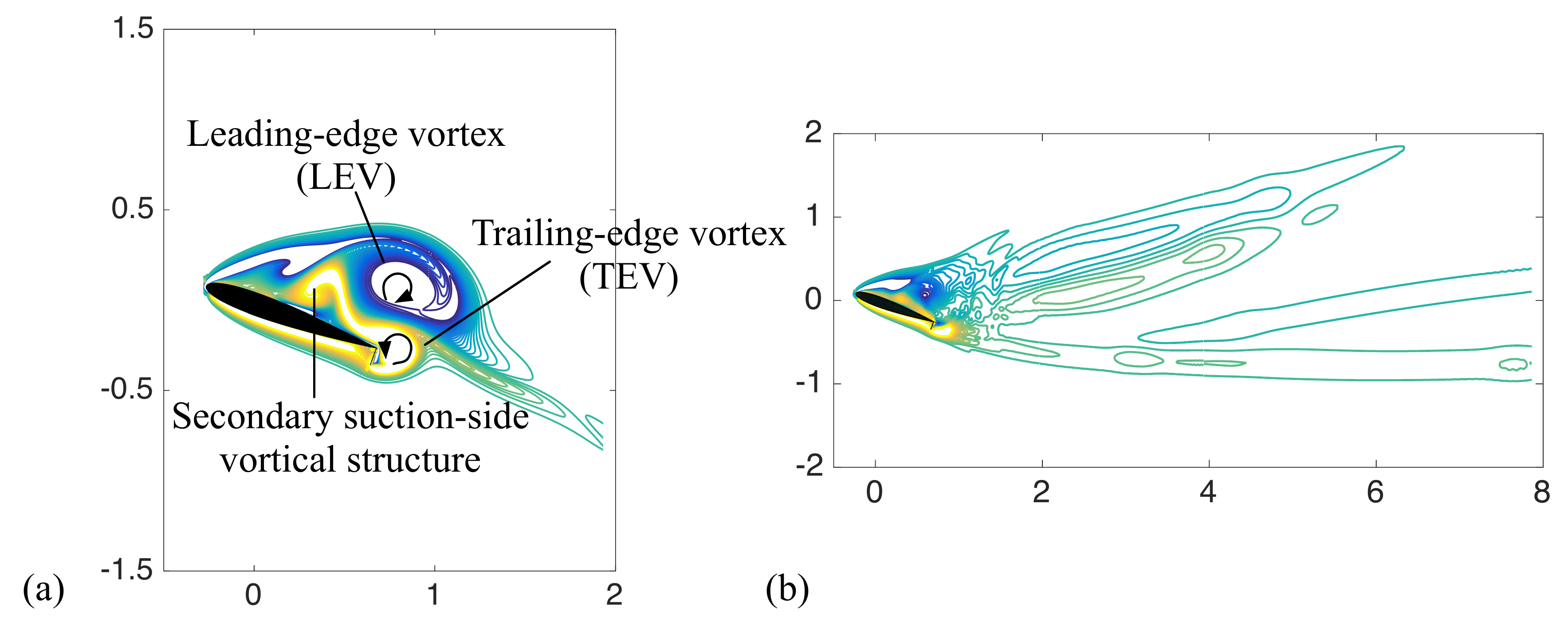}
\caption{Near-wake vortices for NACA 0012 with a Gurney flap of $h/c = 0.1$ at $\alpha=20^{\circ}$ (2P wake regime): (a) instantaneous vorticity field with (1) leading-edge vortex (LEV), (2) trailing-edge vortex (TEV) and (3) secondary suction-side vortical structure, (b) time-averaged vorticity field.}
\label{f:wakereg}
\end{center}
\end{figure}

\begin{figure}[t!]
\begin{center}
\includegraphics[width=1\textwidth]{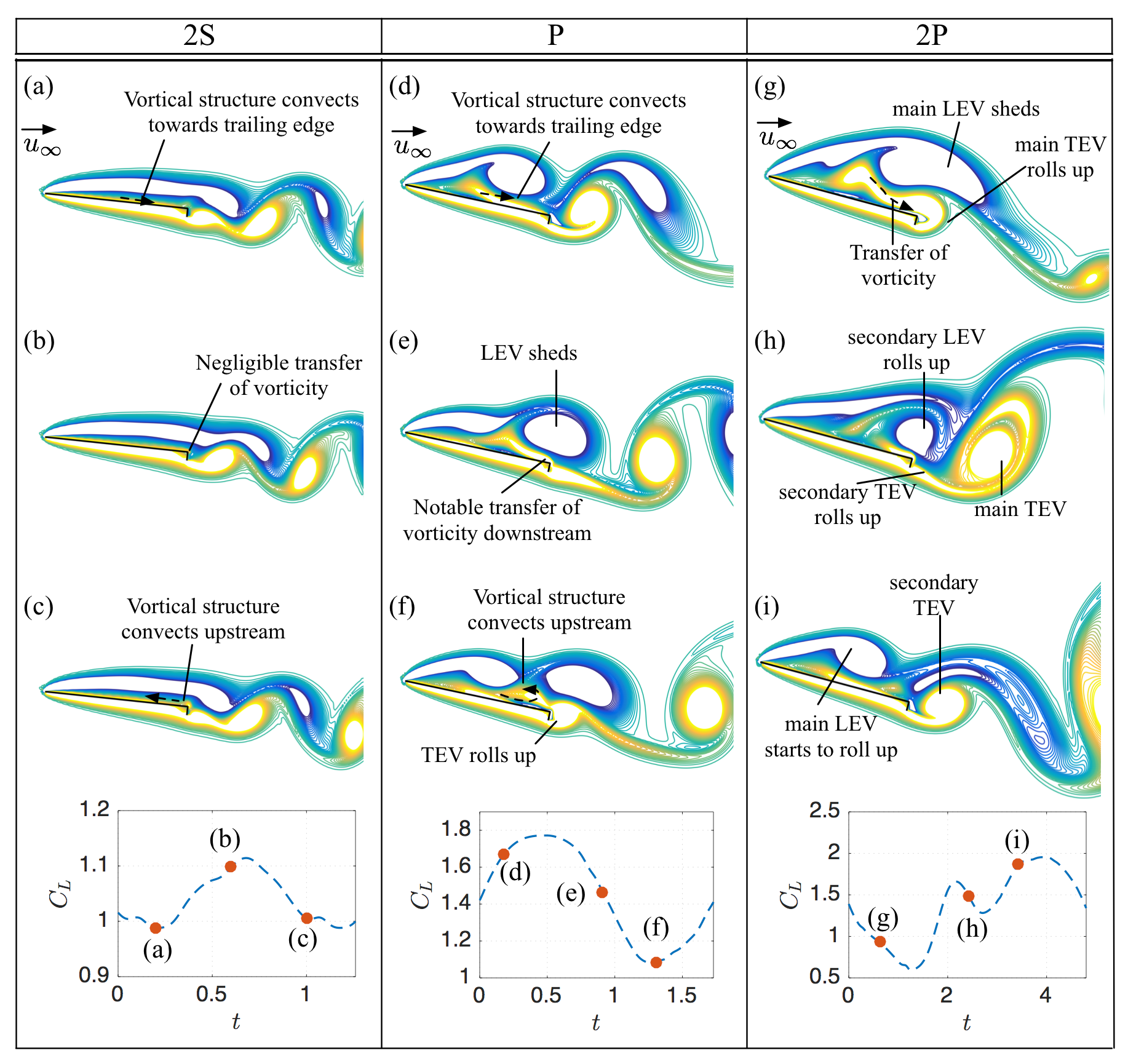}
\caption{Illustration of near field flow of three cases classified under the 2S, (a) - (c), P, (d) - (f) and 2P, (g) - (i) regimes. The figure depicts instantaneous vorticity contour plots of flow over NACA 0000 airfoil at $\alpha = 6^{\circ}, 12^{\circ}$ and $15^{\circ}$ with a Gurney flap of $h/c = 0.06$. Corresponding instantaneous lift data, represented by the \textcolor{RedOrange}{\large$\bullet$}, is also displayed on the bottom.}
\label{f:wakedyn}
\end{center}
\end{figure}

Next, let us analyze the wake dynamics influencing the wake mode transitions, which is responsible for the unsteady aerodynamic forces experienced by the airfoils. We in particular focus on the difference in shear-layer roll-up and near-wake vortical structures that lead to the formation of vortices further downstream. Mainly three regions are examined in detail: suction and pressure sides of the airfoil, and vicinity of the Gurney flap. We depict the canonical near-field wake vortical structures analyzed herein in Fig.~\ref{f:wakereg}.

The main vortical structures observed at the near wake of the airfoil at different angles of attack and Gurney flap heights comprise of the main clockwise rotating leading-edge vortex (LEV) on the suction side and the main counter-clockwise rotating trailing-edge vortex (TEV) on the pressure side, visualized in Fig.~\ref{f:wakereg}(a). Secondary LEV and TEV also shed from the airfoil at high angles of attack or with a large Gurney flap (in the 2P regime). Apart from these vortical structures, the formation of the secondary suction-side vortical structure with positive vorticity between the leading-edge separation and suction-side wall, as shown in Fig.~\ref{f:wakereg}(a), is observed to be important for the wake dynamics. The formation of this secondary structure is a result of the roll-up of the large LEV caused by the separated flow. The formation and behavior of the structures vary with transition of the wake regime.

Let us present a brief summary of the near-wake dynamics, followed by specific details for each wake mode. The formation and periodic shedding of LEV and TEV are expected at moderate angles of attack for flow past an airfoil. Once a sufficiently strong LEV is formed, the secondary suction-side vortical structure emerges on the suction surface. The increase of circulation of the secondary vortical structure with substantial increase of strength of the LEV is observed. Moreover, the strong influence of the LEV and the secondary vortical structure at higher $\alpha$ and $h/c$ (in the 2P regime) causes the TEV to roll-up onto the suction side, eventually leading to transfer of vorticity from the secondary vortical structure to the TEV, as depicted in Fig.~\ref{f:wakereg}(a). Note that the Gurney flap supports this merger by locally delaying the incoming flow as the TEV forms and merges with the secondary suction-side vortical structure. This phenomenon is observed only at very high angles of attack or with a substantially large Gurney flap, particularly in the 2P wake regime. The process results in a broader wake, leading to increase in drag. Moreover, the mean flow is displaced upward with respect to the center of the wake, which is evident for the 2P mode from the time-averaged vorticity field in Fig.~\ref{f:wakereg}(b), decreasing the lift enhancement.

For the 2S regime, Fig.~\ref{f:wakedyn} (a)-(c), we observe the roll-up and convection of a vortical structure with positive vorticity between the leading-edge separation and the suction-side wall, visible at the trailing edge in Fig.~5(c). We refer to this vortical structure as the secondary suction-side vortical structure for the 2S regime. The circulation of the secondary suction-side vortical structure is small and is not strong enough to influence the TEV. The LEV and TEV shed periodically as single vortices, and there is no upward shift of the flow above the center of the wake. These are the reasons for the high lift-to-drag ratios, $\overline{C_L/C_D}$, observed for cases in the 2S wake mode, which are further described in section \ref{sec:aero_forces}.

As the flow transitions to the P regime, Fig.~\ref{f:wakedyn} (d)-(f), the circulation of the near-wake vortical structures increase. The strengthened LEV and secondary suction-side vortical structure influences the TEV to roll up along the outer walls of the Gurney flap with considerable transfer of vorticity from the secondary structure to the TEV. The LEV and TEV shed periodically but a distinct vortex pair is formed compared to the individual vortices in the wake of the 2S regime. The wake width also increases and the flow shifts above the center of the wake. This leads to the increase in drag, decrease in lift enhancement and fall of the $\overline{C_L/C_D}$ curve, as further discussed in section \ref{sec:aero_forces}, for cases in the P regime.

In the 2P regime, Fig.~\ref{f:wakedyn} (g)-(i), the circulation of the secondary suction-side vortical structure is sizable enough to influence the TEV to roll-up onto the suction side, and the vorticity from the secondary vortical structure is transferred to the TEV. The transfer and accumulation of positive vorticity on the suction-side causes the main LEV to shed, followed by the roll-up of the leading-edge shear layer to form the second LEV. The main TEV sheds after considerable build-up of circulation due to the transfer of vorticity from the secondary vortical structure. This leads to roll-up of the second TEV on the pressure side. By the time the second LEV sheds from the trailing edge, the second TEV rolls-up along the outer walls of the Gurney flap. The influence from the secondary vortical structure is not strong enough for the roll-up of the second TEV onto the suction side. The second LEV convects downward pairing up with the secondary TEV about $10c$ downstream. The wake width is increased considerably compared to the other unsteady regimes, with the flow shifting above and below the center of the wake. This wake dynamics lead to increase of drag, decrease of lift enhancement and degradation of the $\overline{C_L/C_D}$ curve, which are described in section \ref{sec:aero_forces}.


\subsection{Aerodynamic Forces}\label{sec:aero_forces}

\begin{figure}[t!]
 \begin{subfigmatrix}{2}
  \subfigure[NACA 0000]{\includegraphics{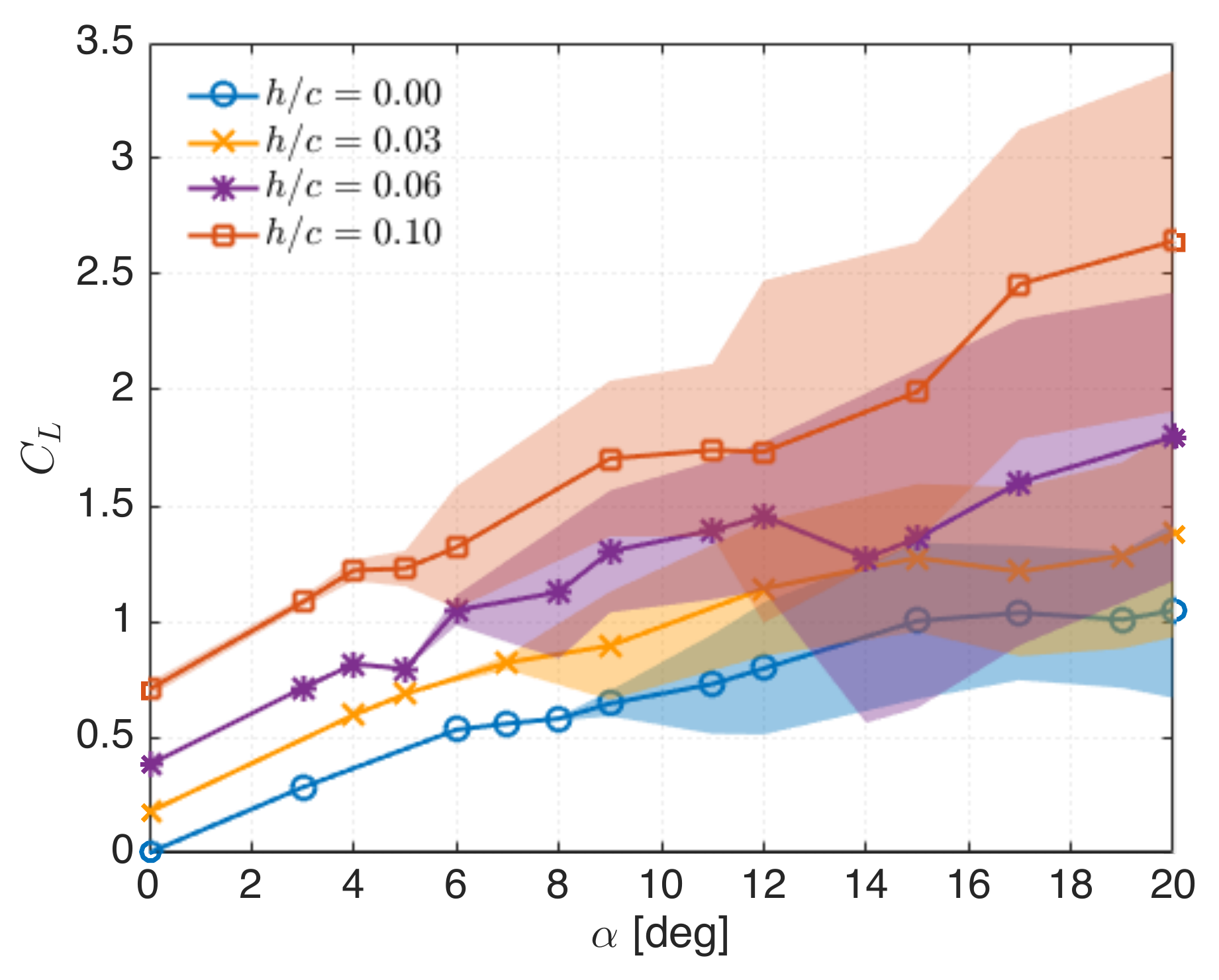}}
  \subfigure[NACA 0006]{\includegraphics{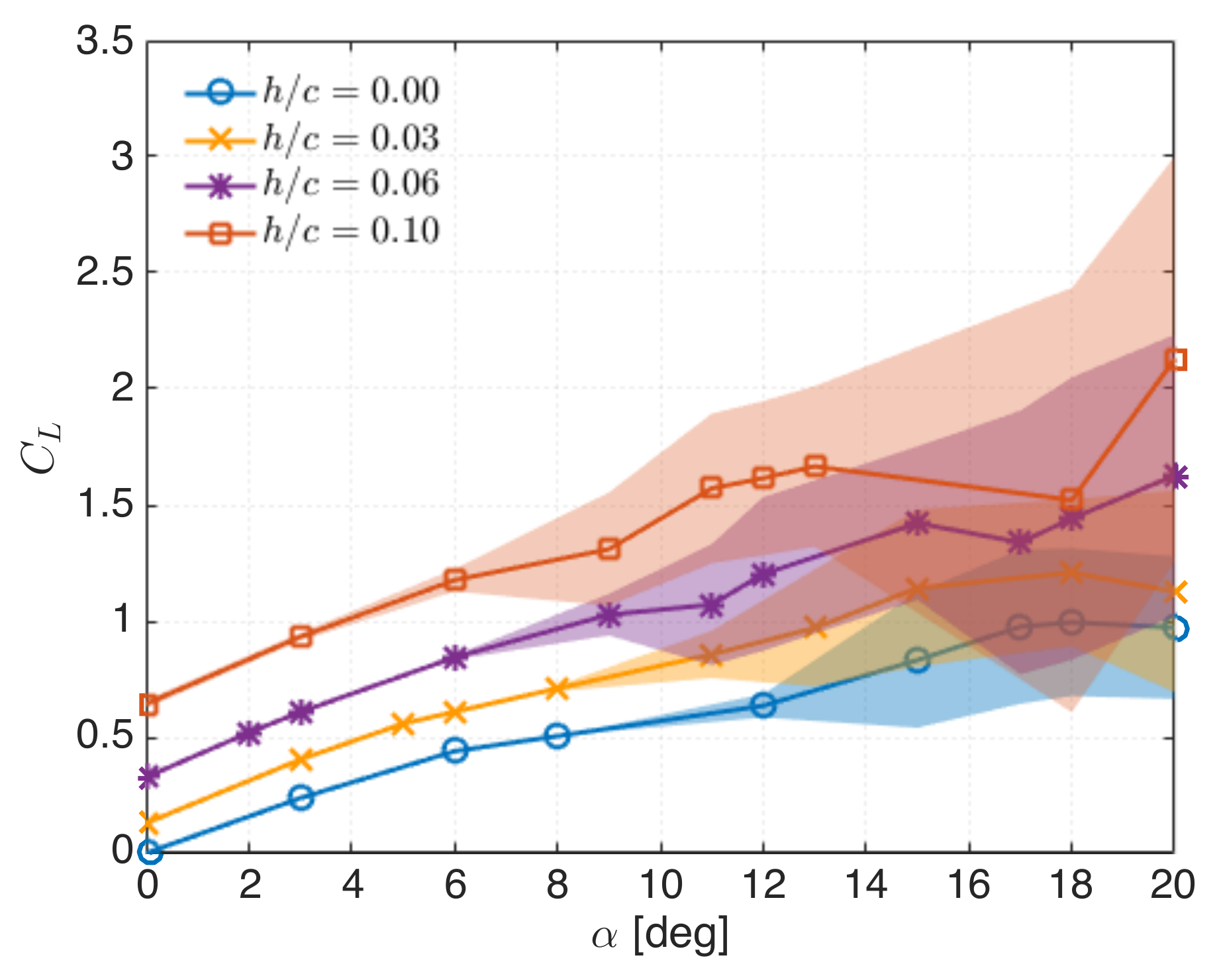}}
  \subfigure[NACA 0012]{\includegraphics{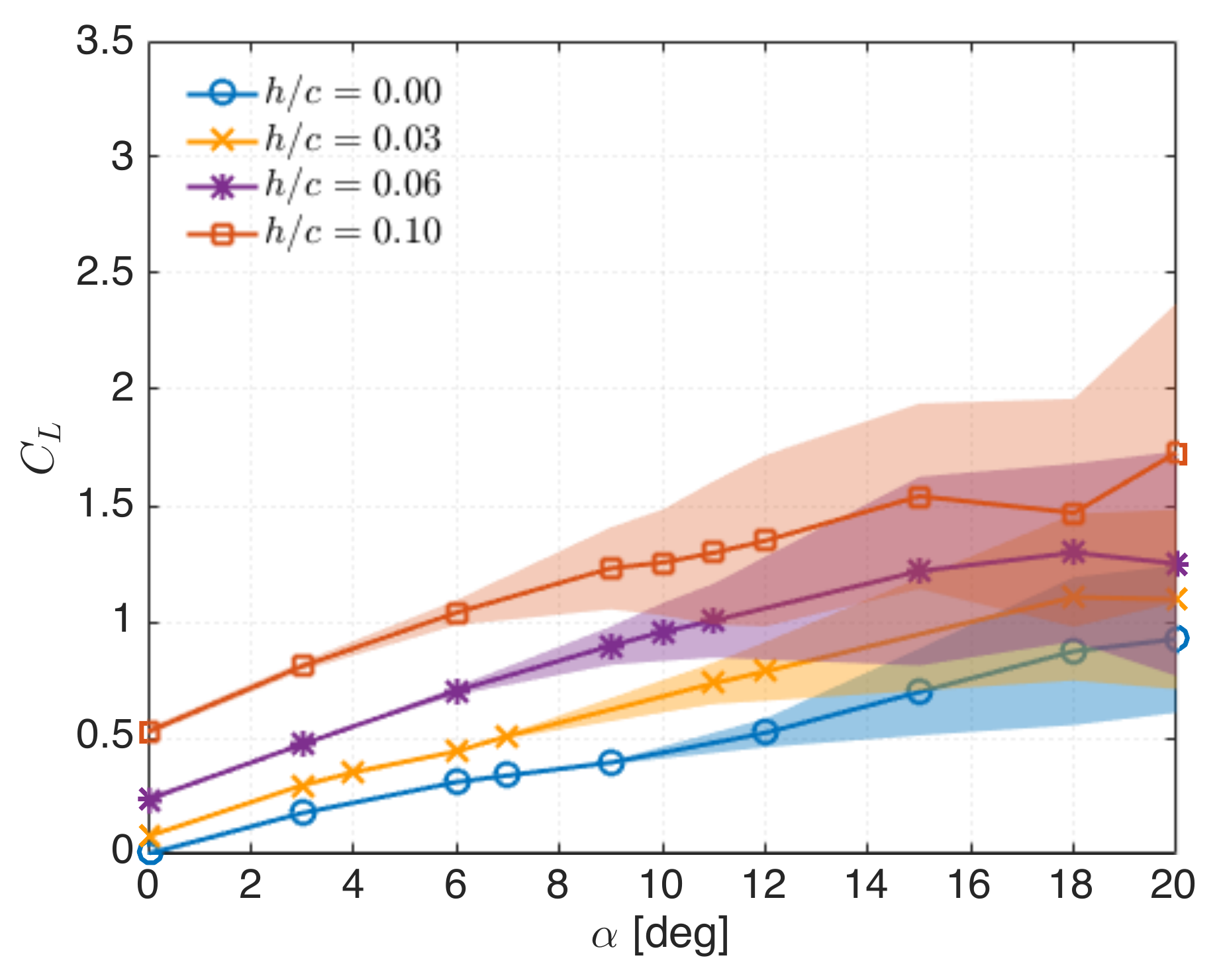}}
  \subfigure[NACA 0018]{\includegraphics{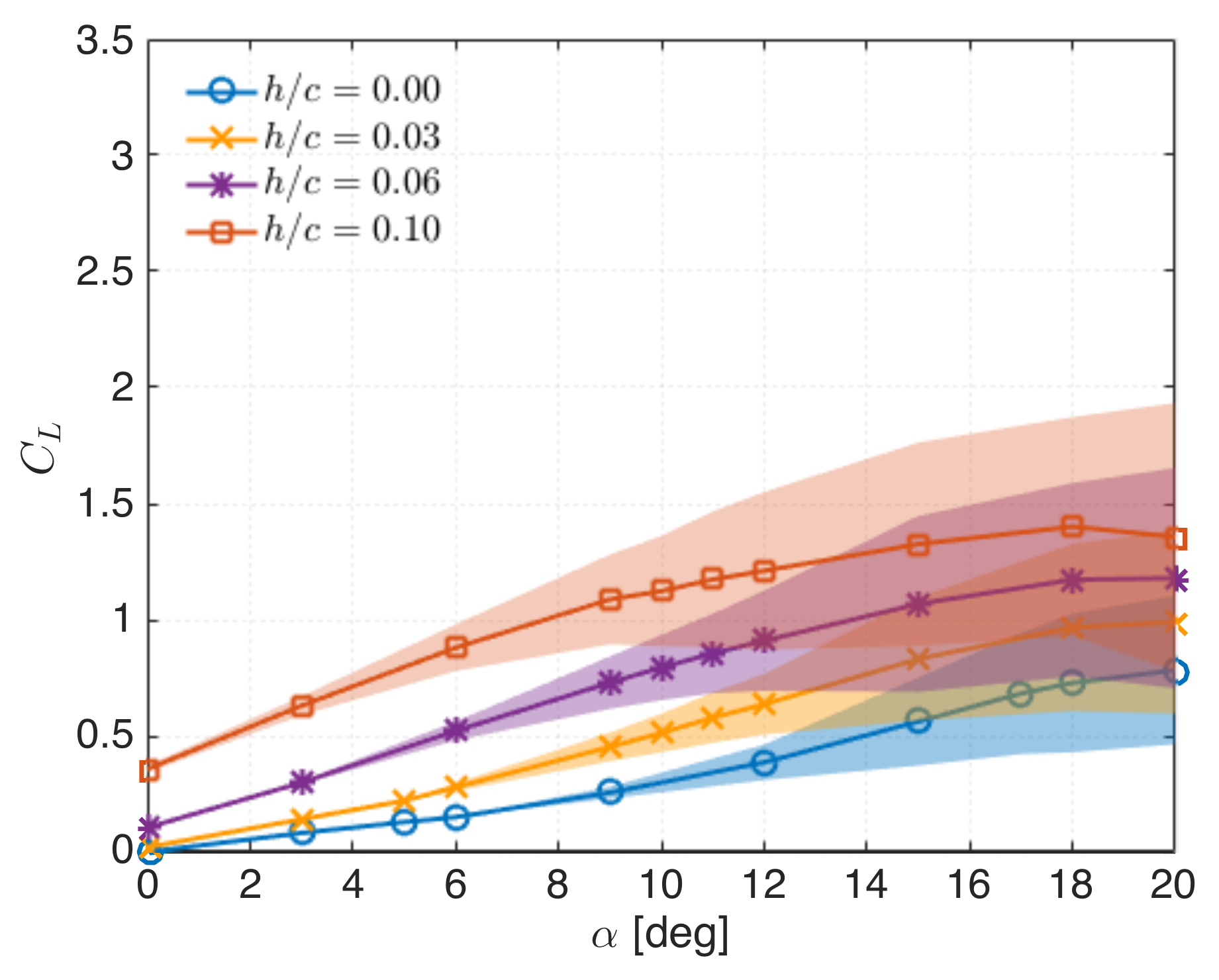}}
 \end{subfigmatrix}
  \caption{Lift coefficient $C_L$ over angle of attack $\alpha$ for NACA 0000, 0006, 0012 and 0018 airfoils with select Gurney flap heights. The time-averaged $\overline{C_L}$ values are represented by the solid lines and fluctuations in $C_L$ are represented by the shaded region for each configuration.}
 \label{f:CL}
\end{figure}

Next, we examine the effects of the wake modifications on the aerodynamic forces from the use of Gurney flaps. The main objective of attaching a Gurney flap to the trailing edge of an airfoil is to enhance the lift force experienced by the airfoil. Current results show that at a low Reynolds number of $Re = 1000$, the Gurney flap is able to generate high levels of lift forces. It is also seen that as the height of the flap is increased, lift also increases. The lift coefficient, $C_L$, of representative cases for each airfoil is shown in Fig.~\ref{f:CL}, where the time-averaged values, $\overline{C_L}$, are depicted by the solid lines and the fluctuations in the forces are represented by the shaded regions. The blue lines for $h/c = 0.00$ cases depict the baseline results without a Gurney flap for all the airfoils.

As discussed above, $\overline{C_L}$ increases with $\alpha$ and $h/c$ values for all airfoils. Lift force enhancement of more than twice the baseline values is achieved using Gurney flap, particularly at $h/c = 0.10$ for all airfoils in almost all cases. Due to the presence of the Gurney flap, effective camber of the airfoil increases, attributing to lift enhancement. Fluctuations in the lift forces appear when the flow transitions from steady to unsteady state (2S regime). For a given angle of attack, $\overline{C_L}$ as well as the fluctuations of $C_L$ are amplified with increase in Gurney flap height. 

At higher angles of attack, sudden jumps or bursts in $C_L$ fluctuations are observed for all the cases. The fluctuations in $C_L$ exhibits trends of sudden increase in magnitude with increase in $\alpha$, suggesting a transition from one flow regime to another with different Gurney flap heights and angles of attack. The transitions correspond to the wake mode shift from the 2S mode to the P mode, and then to the 2P mode. Thus, the airfoil experiences the highest magnitude of $\overline{C_L}$ as well as the fluctuations of $C_L$ when the wake is classified under the 2P mode. Again, the transition of the wake from the steady regime through the 2P regime can be evidently observed by the jagged growth of $C_L$ fluctuations with $\alpha$ for the NACA 0000 airfoil. The transition of the wake from the P to the 2P mode is clearly observed for the NACA 0006 and 0012 cases at high angles of attack.

\begin{figure}[t!]
 \begin{subfigmatrix}{2}
  \subfigure[NACA 0000]{\includegraphics{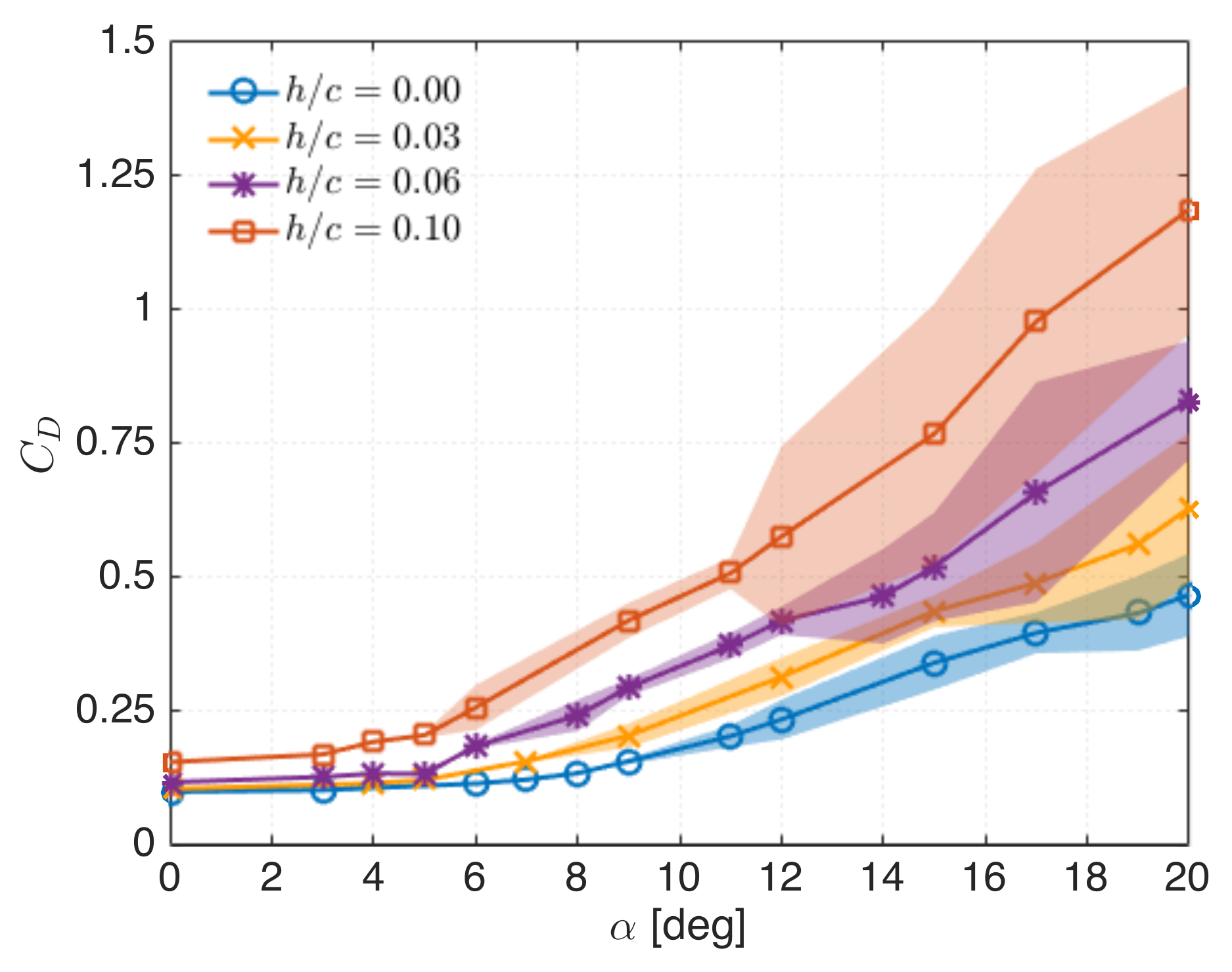}}
  \subfigure[NACA 0006]{\includegraphics{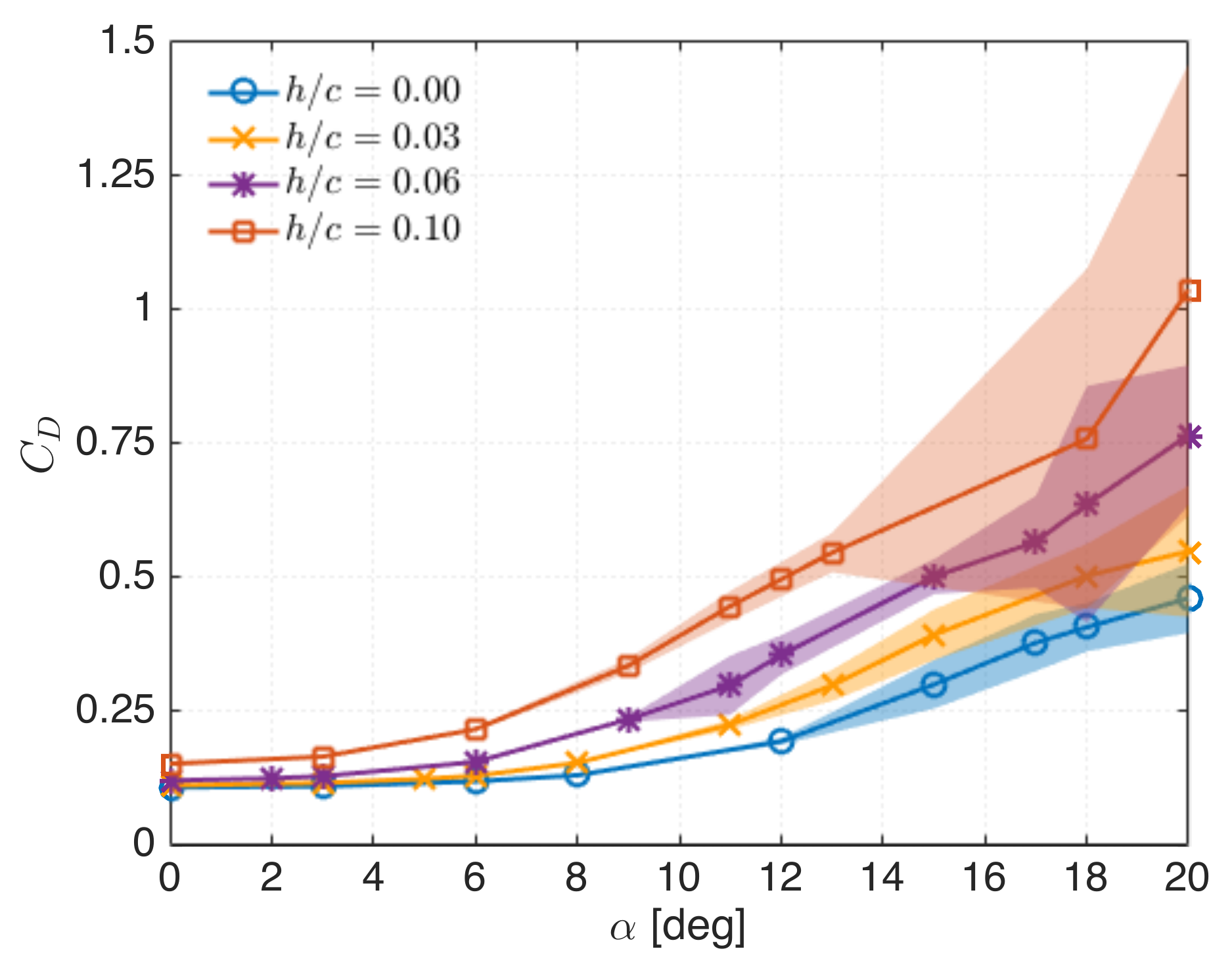}}
  \subfigure[NACA 0012]{\includegraphics{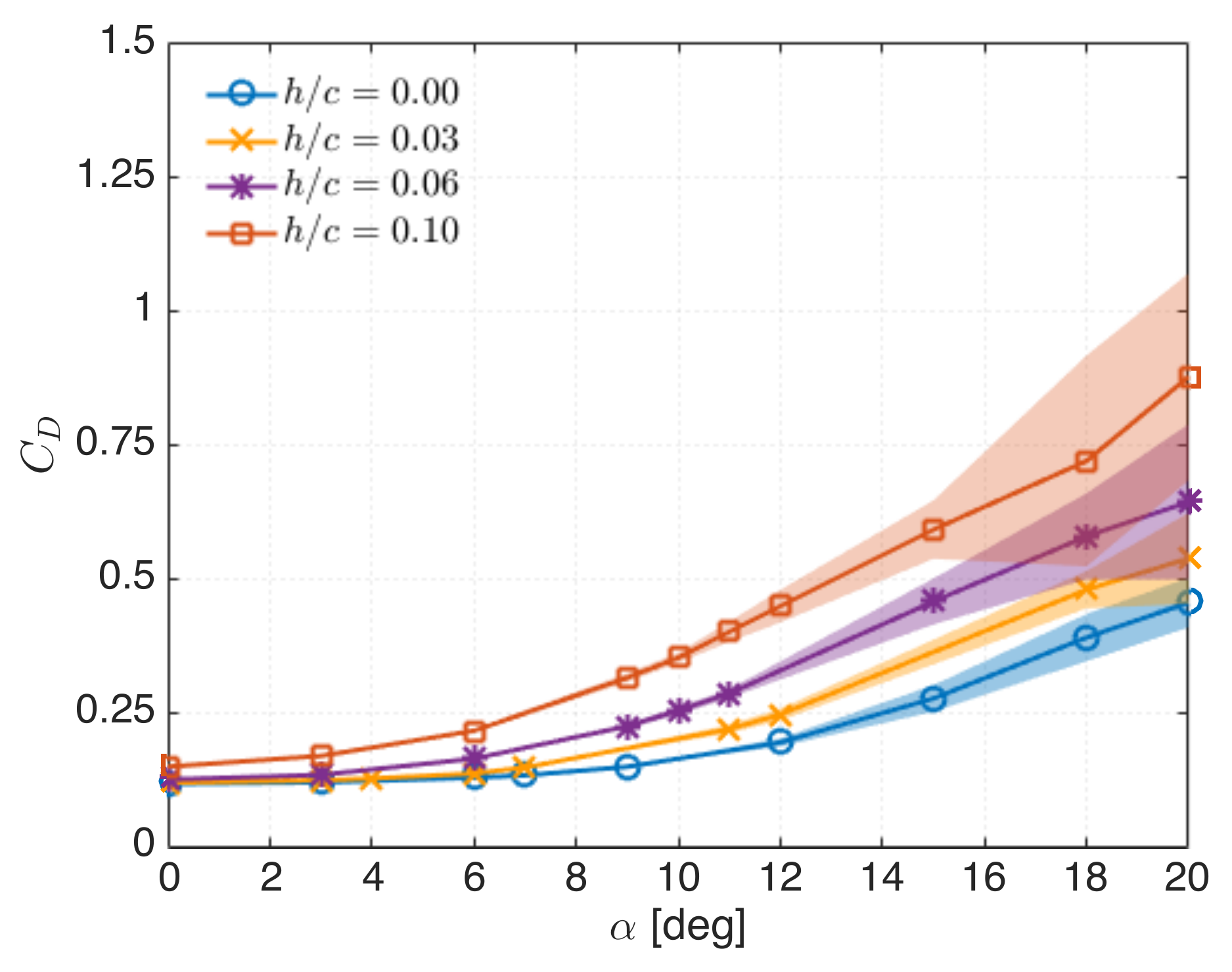}}
  \subfigure[NACA 0018]{\includegraphics{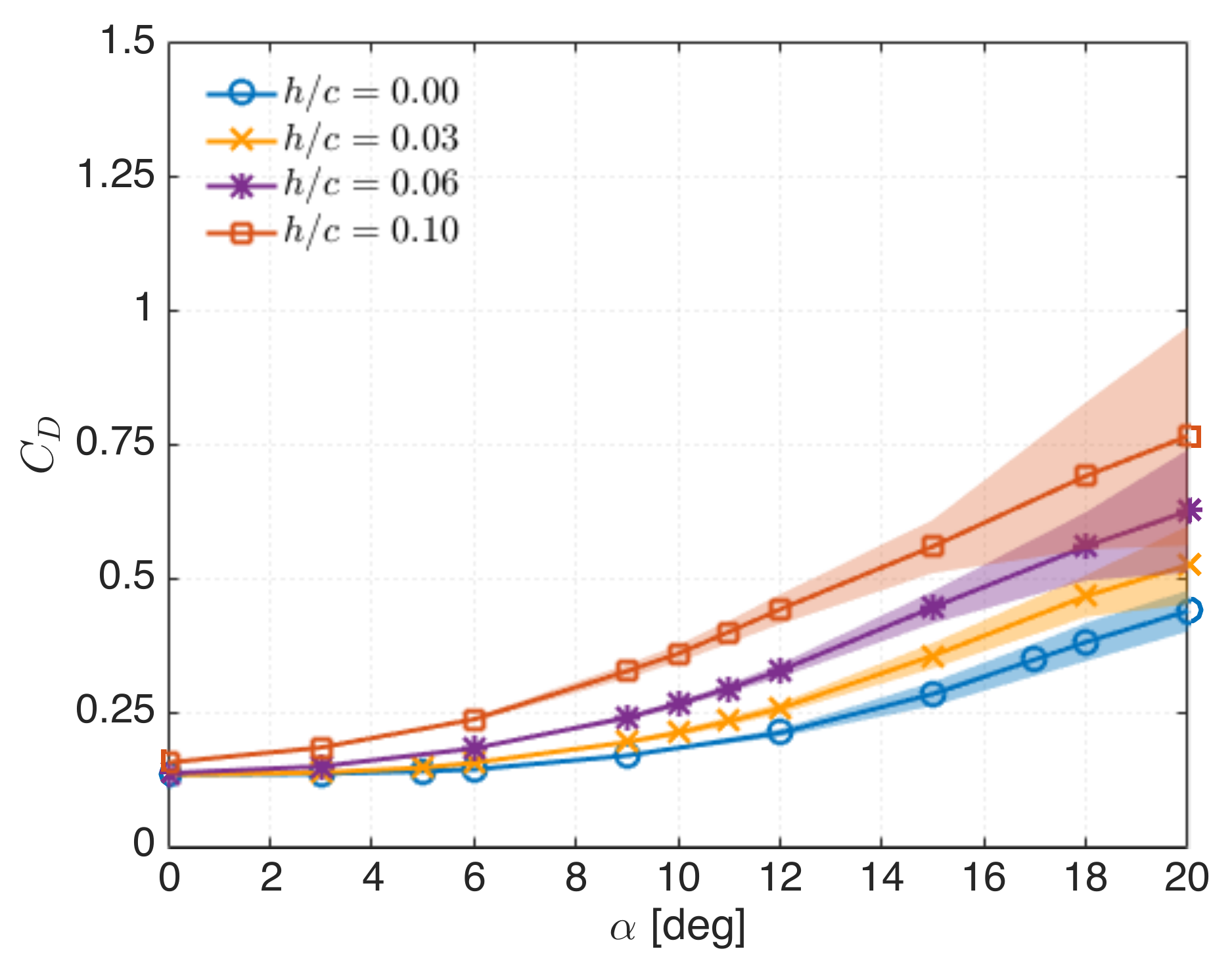}}
 \end{subfigmatrix}
  \caption{Drag coefficient $C_D$ over angle of attack $\alpha$ for NACA 0000, 0006, 0012 and 0018 airfoils with select Gurney flap heights. The time-averaged $\overline{C_D}$ values are represented by the solid lines and fluctuations in $C_D$ are represented by the shaded region for each configuration.}
 \label{f:CD}
\end{figure}

The benefit of lift enhancement is however accompanied with a penalty of some drag increase. The drag coefficients, $C_D$, of representative cases for each airfoil are shown in Fig.~\ref{f:CD}. As it can be observed, $\overline{C_D}$ increases with $\alpha$ and $h/c$ values. The $\overline{C_D}$ for all cases have low values at lower angles of attack. The appearance of fluctuations in $C_D$ in conjunction with the wake transition from the steady regime to the 2S regime is also observed at lower $\alpha$, but the magnitudes of the fluctuations are small. Also, the sudden and jagged variations in the fluctuations of $C_D$ with increase in $\alpha$ are small compared to that observed for the $C_L$ fluctuations. The fluctuations in $C_D$ increase suddenly at high angles of attack when the wake transition from the P mode to the 2P mode, also observed above for lift, for all airfoils.

The thickness of the airfoil plays a major role on the aerodynamic forces experienced by the airfoil. With increase in airfoil thickness, the effect of the Gurney flap is overshadowed. As a result, different trends are observed for lift and drag on the airfoils. For the range of angles of attack considered in this study, the magnitudes of $\overline{C_L}$ and fluctuations in $C_L$ decreases for thicker airfoils. Whereas, the drag force increases with airfoil thickness at low angles of attack, which is expected for the baseline cases, but the trend reverses at higher angles of attack and drag generated decreases for thicker airfoils compared to thinner airfoils. Another behavior observed are for the lift and drag slopes. The lift slope, $a = {\rm d}\overline{C_L} / {\rm d} \alpha$ follow a jagged trend for thinner airfoils, whereas the slope is smoother for thicker airfoils. The same is observed for the drag slope, although the fluctuations of the drag slope for thinner airfoils are small compared to that of lift slope. All the above observations further emphasizes the reduced influence from the Gurney flap due to the overshadowing effect of the airfoil thickness.

\begin{figure}[t!]
 \begin{subfigmatrix}{2}
  \subfigure[NACA 0000]{\includegraphics{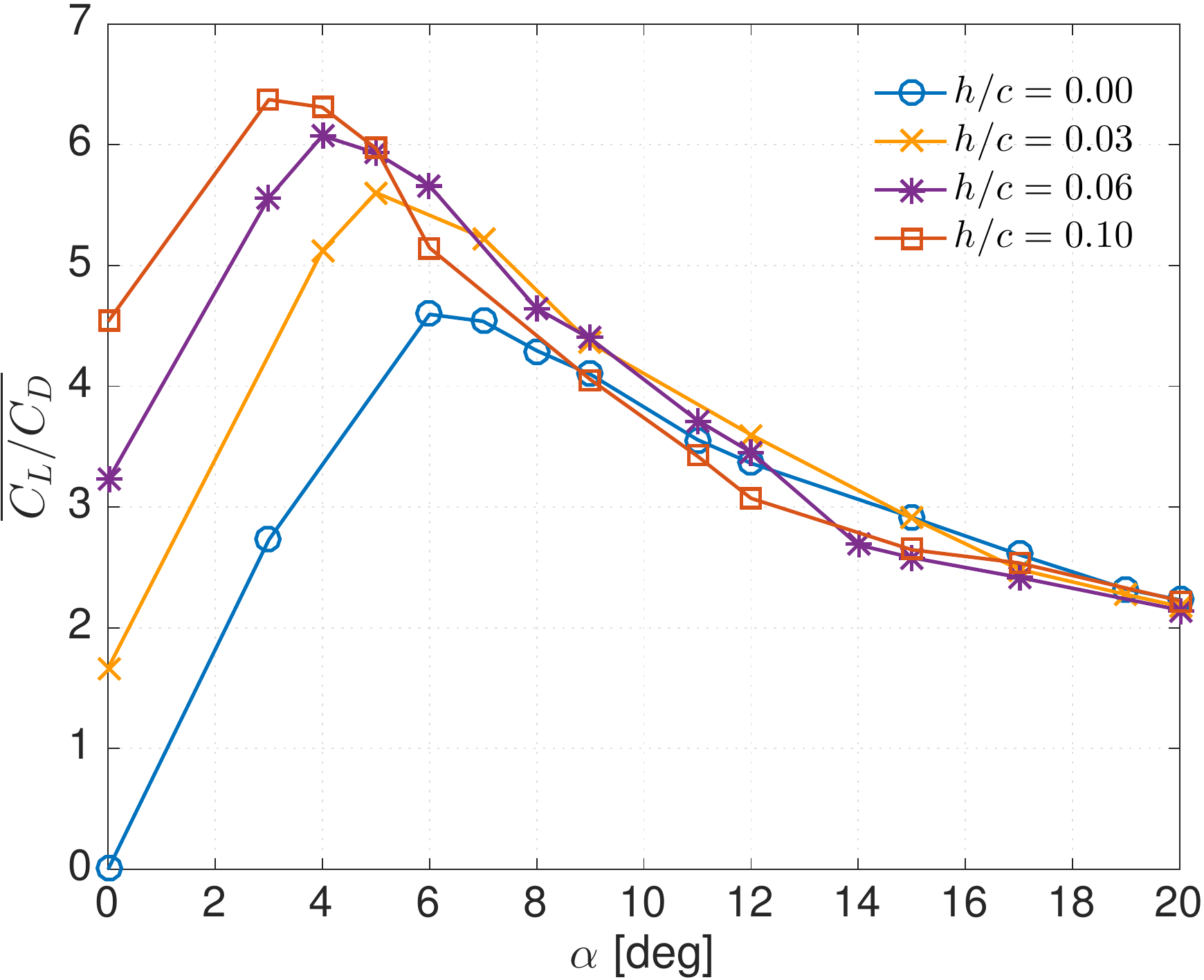}}
  \subfigure[NACA 0006]{\includegraphics{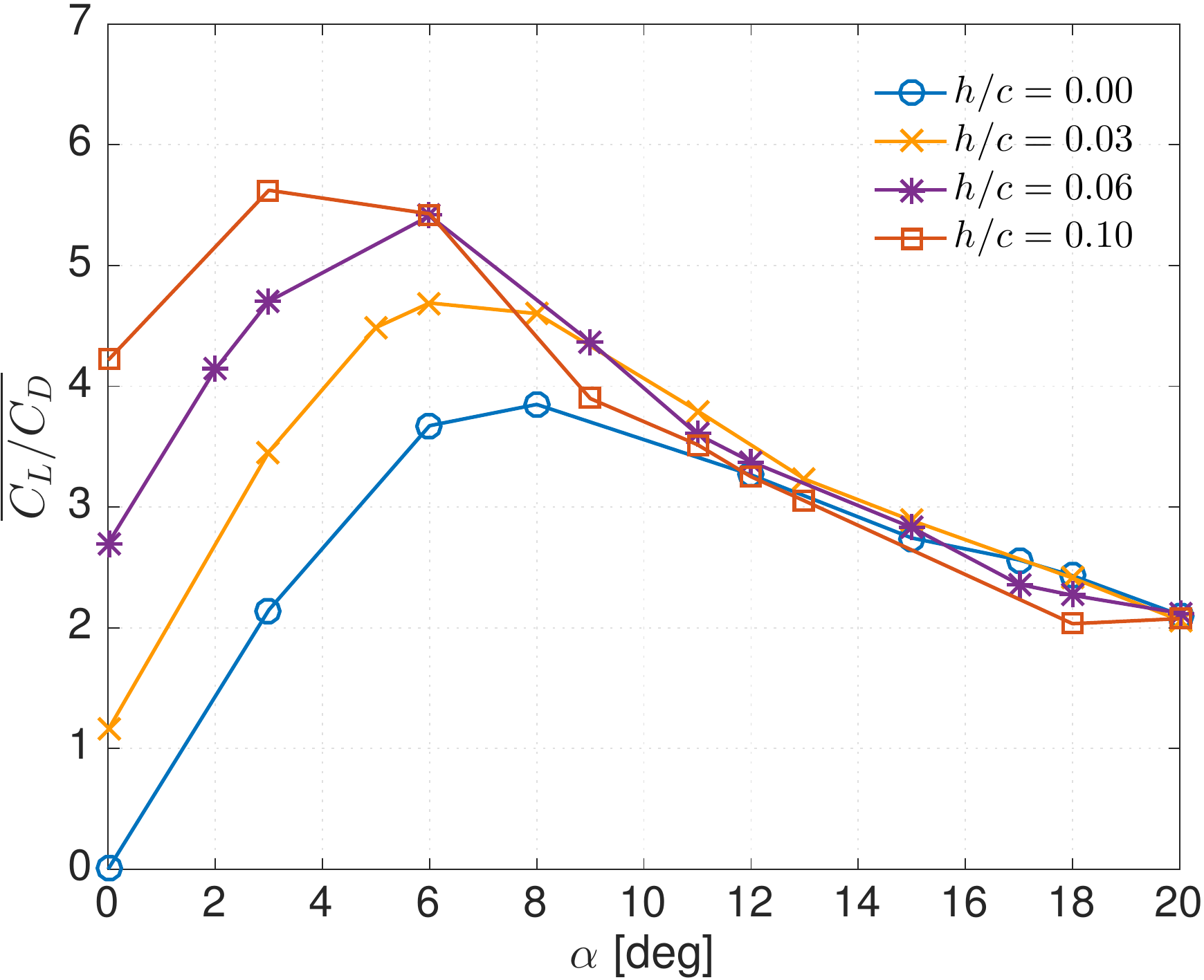}}
  \subfigure[NACA 0012]{\includegraphics{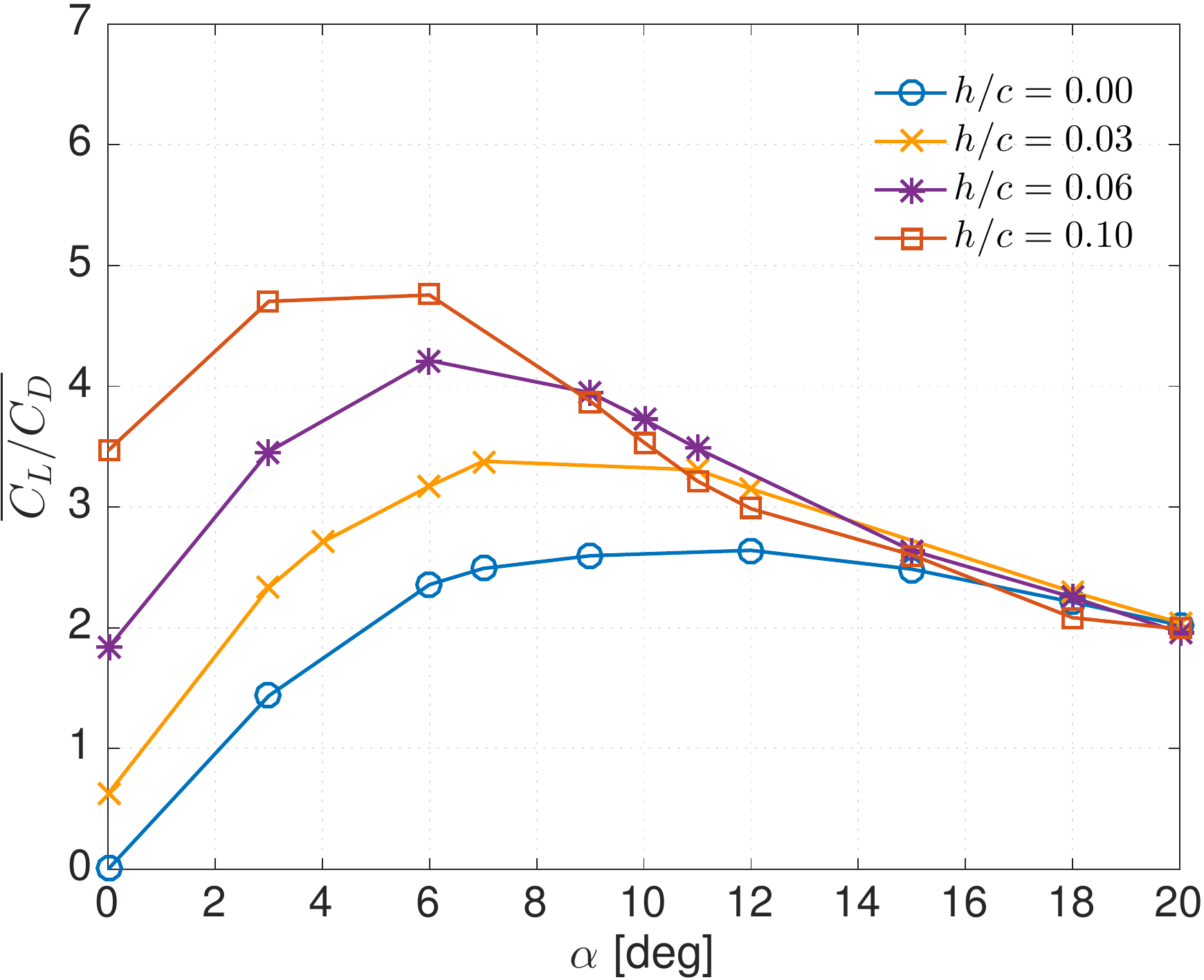}}
  \subfigure[NACA 0018]{\includegraphics{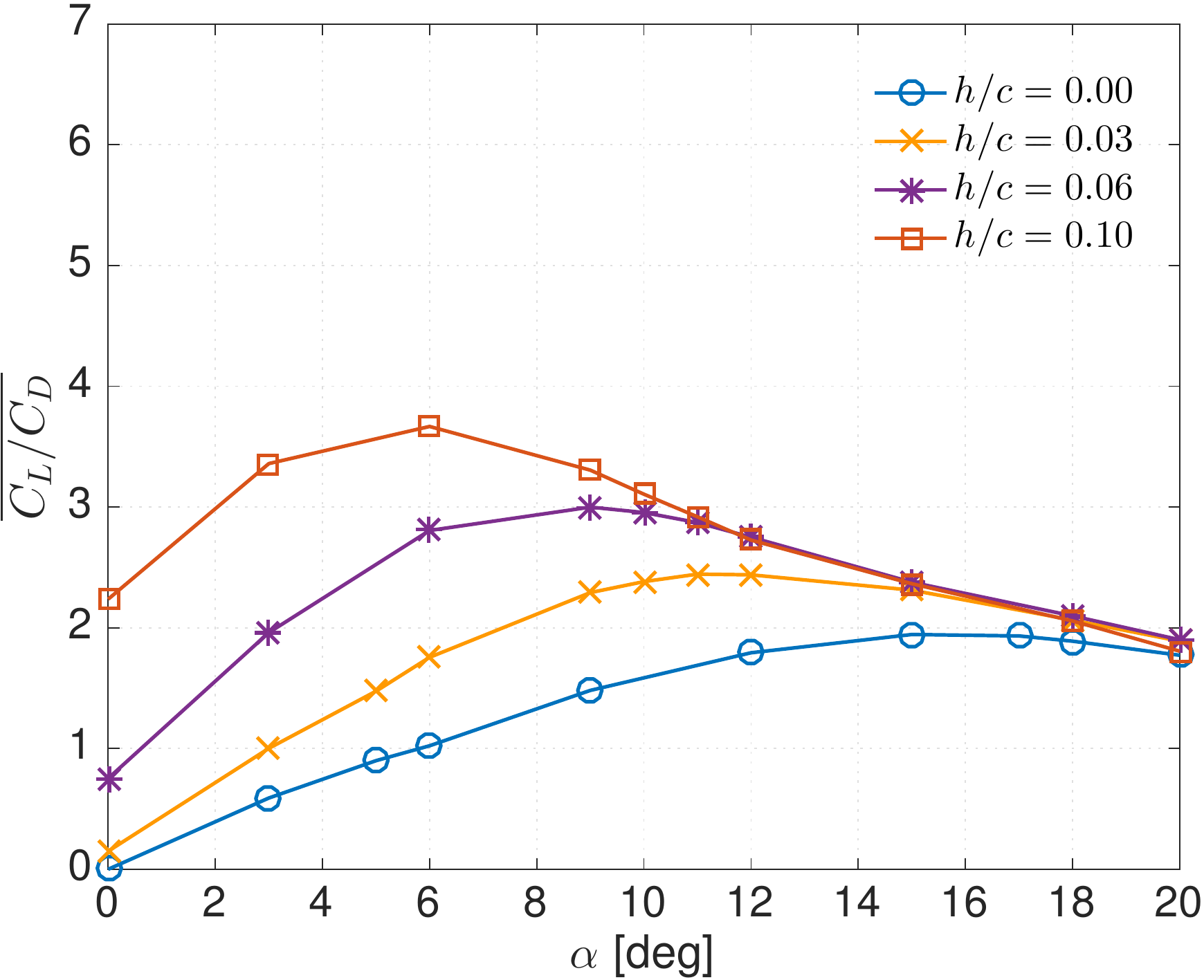}}
 \end{subfigmatrix}
  \caption{Lift-to-drag ratio $\overline{C_L/C_D}$ over angle of attack $\alpha$ for NACA 0000, 0006, 0012 and 0018 airfoils with Gurney flap.}
 \label{f:aero}
\end{figure}

Lift-to-drag ratio, $\overline{C_L/C_D}$, for all airfoils is examined and depicted in Fig.~\ref{f:aero}. With increase in $\alpha$, $\overline{C_L/C_D}$ increases drastically, reaching its maximum value, and then decreases gradually. Significant enhancement of $\overline{C_L/C_D}$ is achieved with the utilization of a Gurney flap. The $\overline{C_L/C_D}$ at lower angles of attack, in the range of $\alpha \in [0^{\circ},12^{\circ}]$, is observed to be higher compared to baseline cases in general for all airfoils and flap heights. This range of $\alpha$ is smaller for thinner airfoils (NACA 0000 and 0006) although the peak value of $\overline{C_L/C_D}$ is higher compared to thicker airfoils (NACA 0012 and 0018). This range of $\alpha$ corresponds to the steady and 2S regime for all the airfoils, where the drag remains low. At higher range of $\alpha$, $\overline{C_L/C_D}$ decreases and reach values close to the baseline values for the controlled cases of all airfoils. These range of $\alpha$ corresponds to the P and 2P regimes of the respective airfoils where the highest magnitude of drag is experienced by the airfoils. The $\overline{C_L/C_D}$ curves degrade below the baseline cases for NACA 0000 and 0006 at higher $\alpha$.\\

All the above observations show that an airfoil with a large Gurney flap at high angles of attack experiences large forces. With increase in $\alpha$, thinner airfoils (NACA 0000 and 0006) experience the highest increase in lift and drag, whereas thicker airfoils (NACA 0012 and 0018) attributes with smoother lift and drag slopes. The enhanced lift-to-drag ratio brought about by the Gurney flap is maintained over a wider range of $\alpha$ and $h/c$ values in thicker airfoils, while thinner airfoils experience higher magnitude of $\overline{C_L/C_D}$ over a narrower range of $\alpha$. Thus, for high levels of lift generation and $\overline{C_L/C_D}$ requirements, thinner airfoils can be suitable amongst the considered airfoils. In contrast, to obtain the enhanced $\overline{C_L/C_D}$ over a wider range of angles of attack and lower drag when Gurney flap is attached, thicker airfoils perform well, although with a relatively lower magnitude of $\overline{C_L/C_D}$.

\section{Analysis of Three-Dimensional Flows}\label{sec:3D}

\begin{figure}[t!]
\begin{center}
  	\begin{tabular}{|c|c|c|c|}
	\hline
      		  & $\alpha = 6^\circ$  & $\alpha = 12^\circ$ & $\alpha = 18^\circ$ \\ \hline \hline  
      	    	\begin{sideways}Top view\end{sideways} & \includegraphics[width=0.31\textwidth]{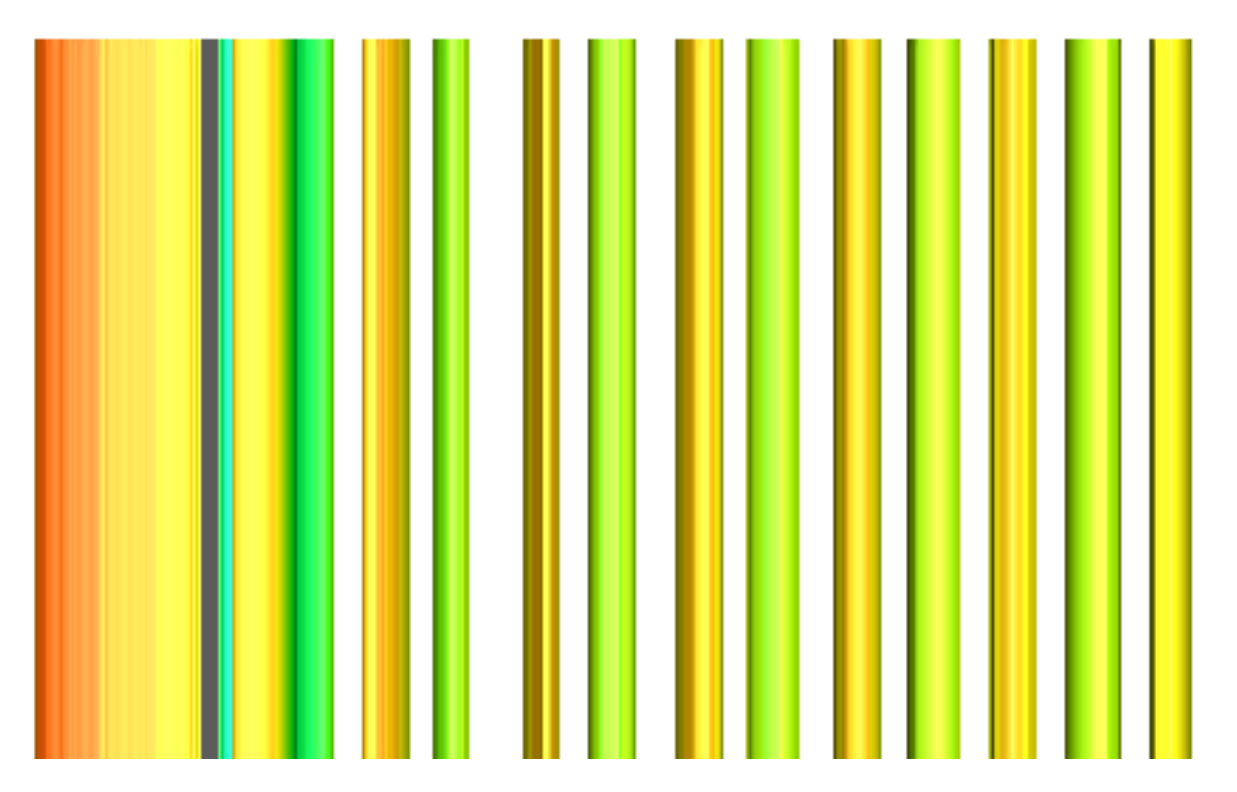} & \includegraphics[width=0.31\textwidth]{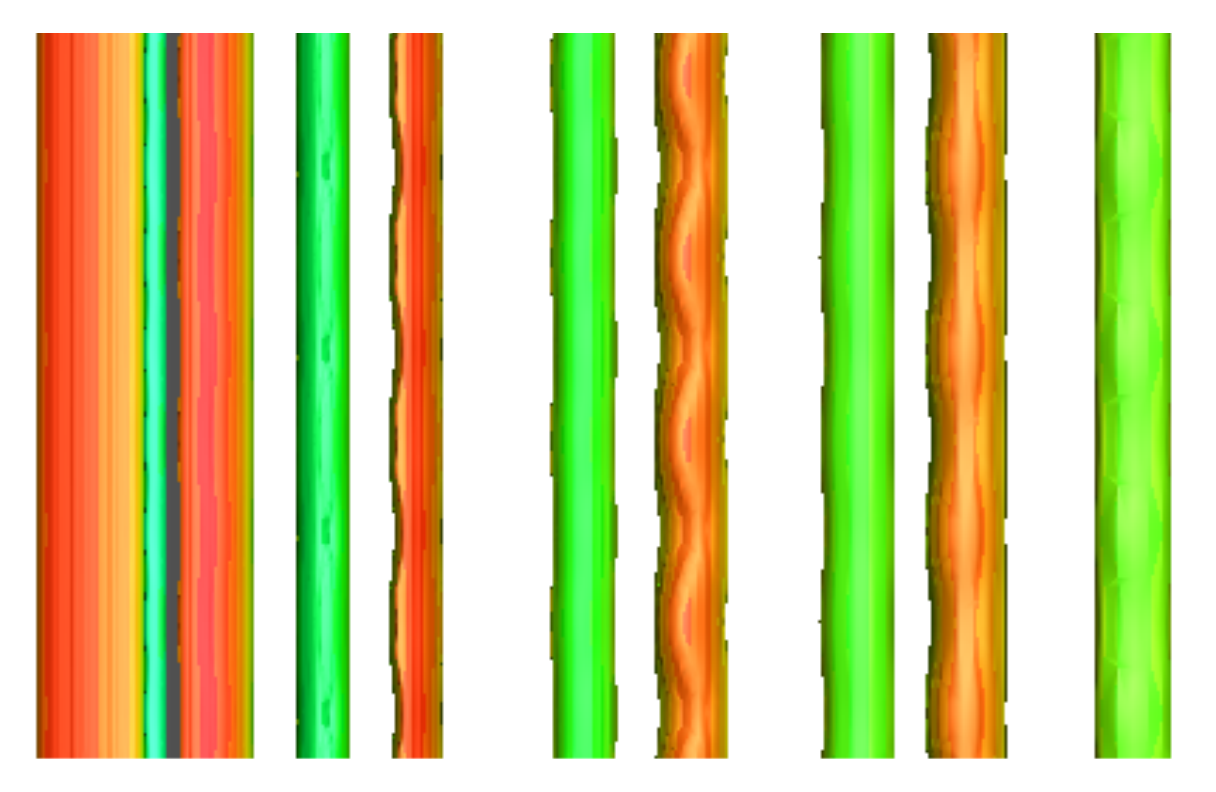} & \includegraphics[width=0.31\textwidth]{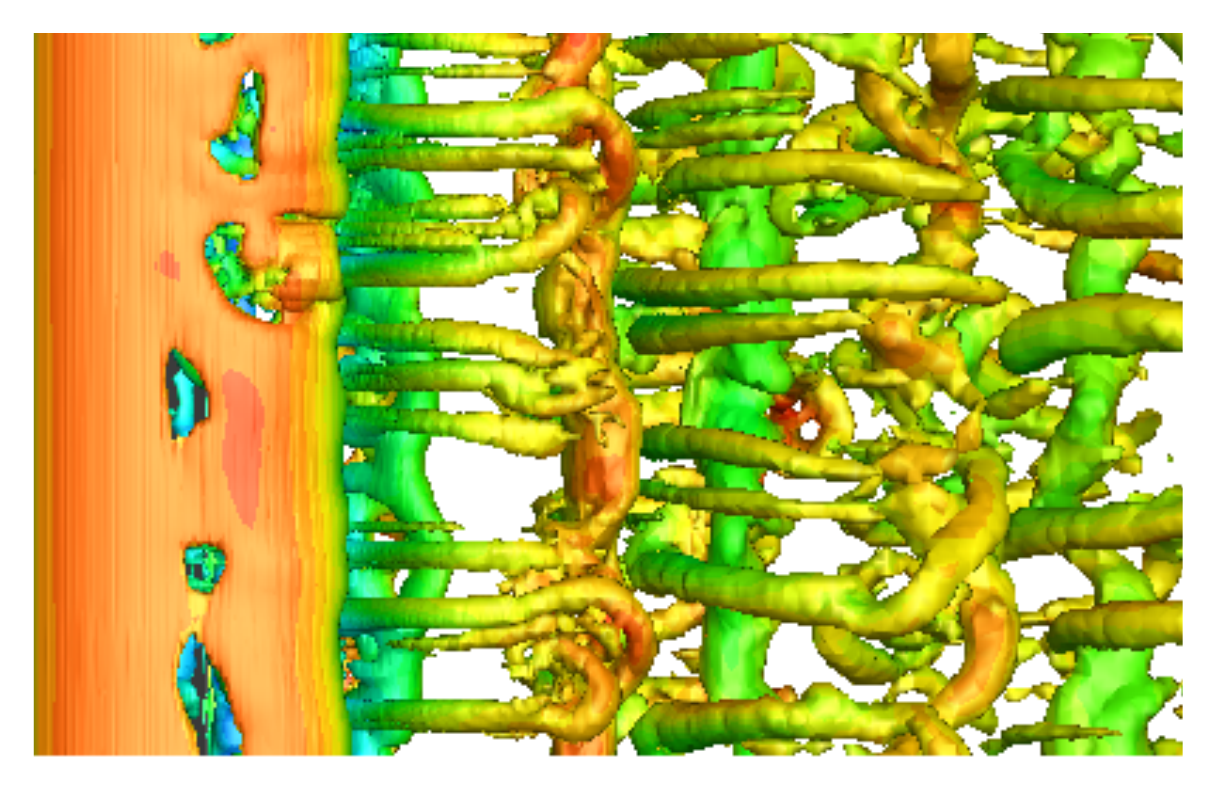} \\ \hline        
          	\begin{sideways}Perspective view\end{sideways} & \includegraphics[width=0.31\textwidth]{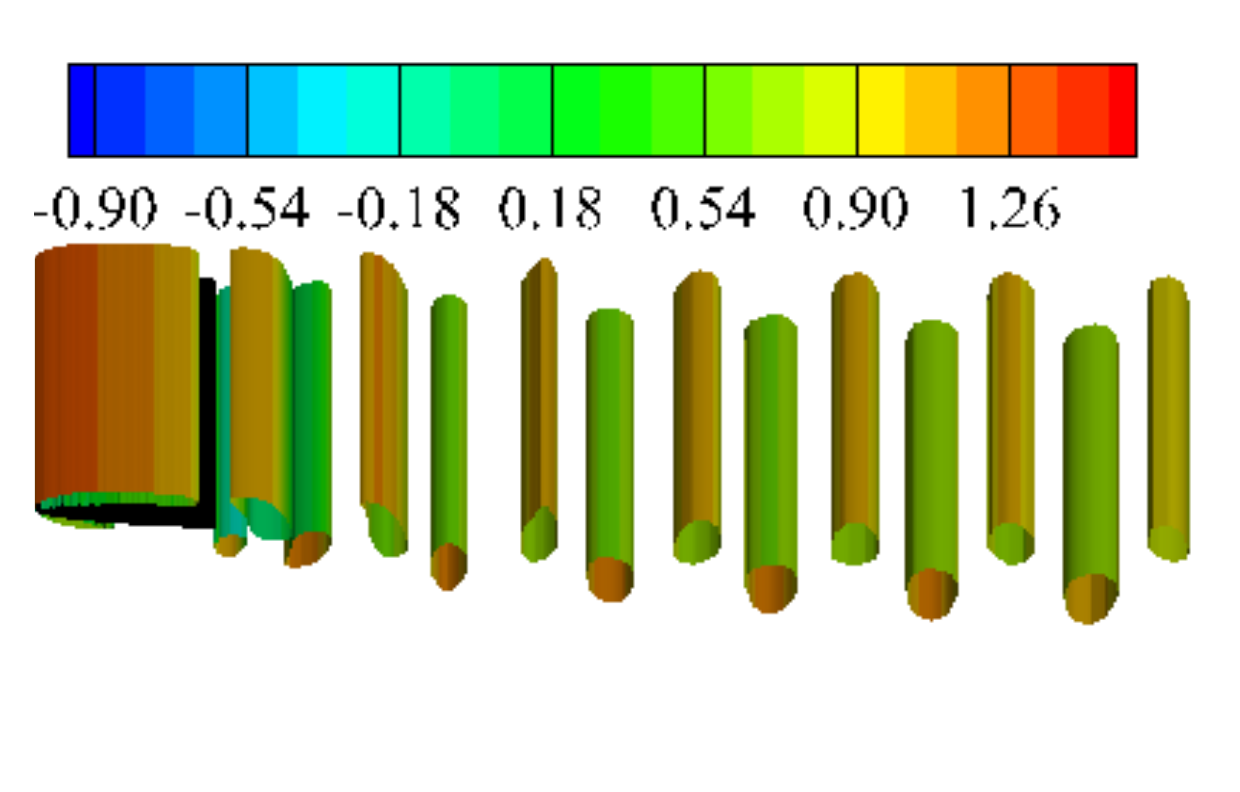} & \includegraphics[width=0.31\textwidth]{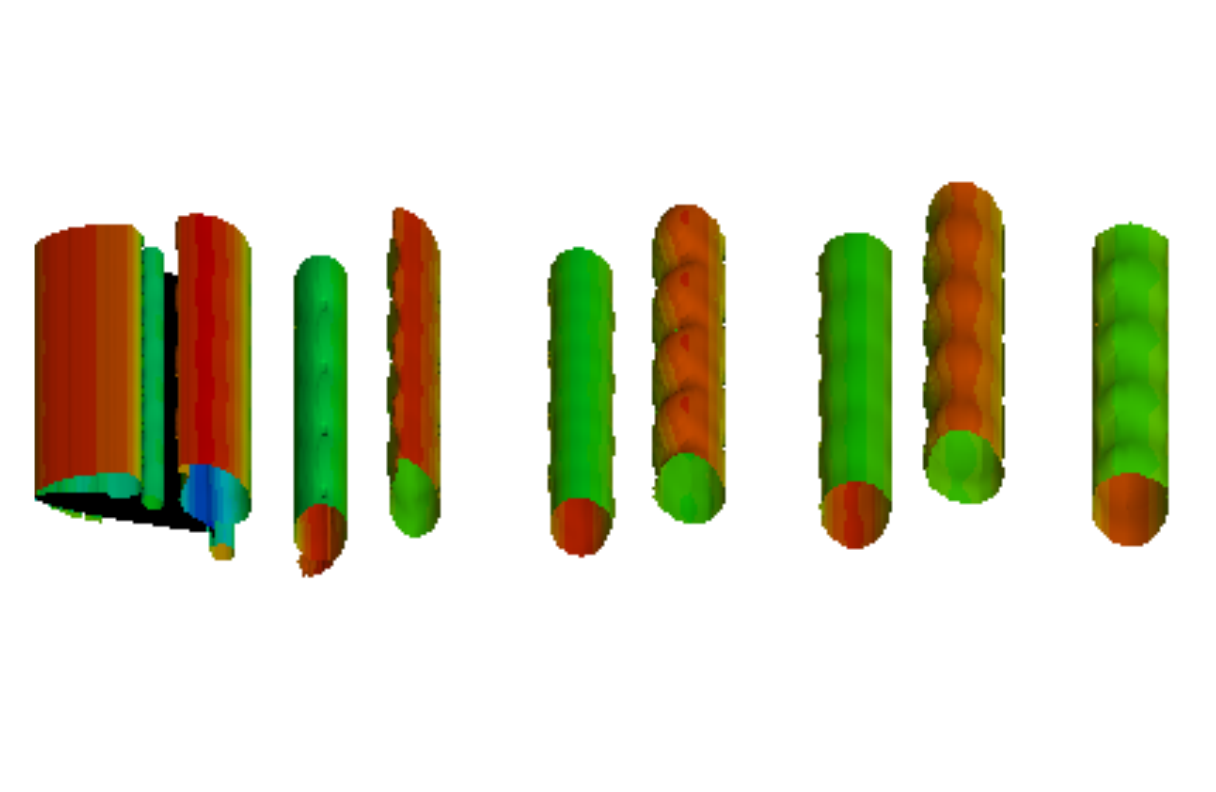} & \includegraphics[width=0.31\textwidth]{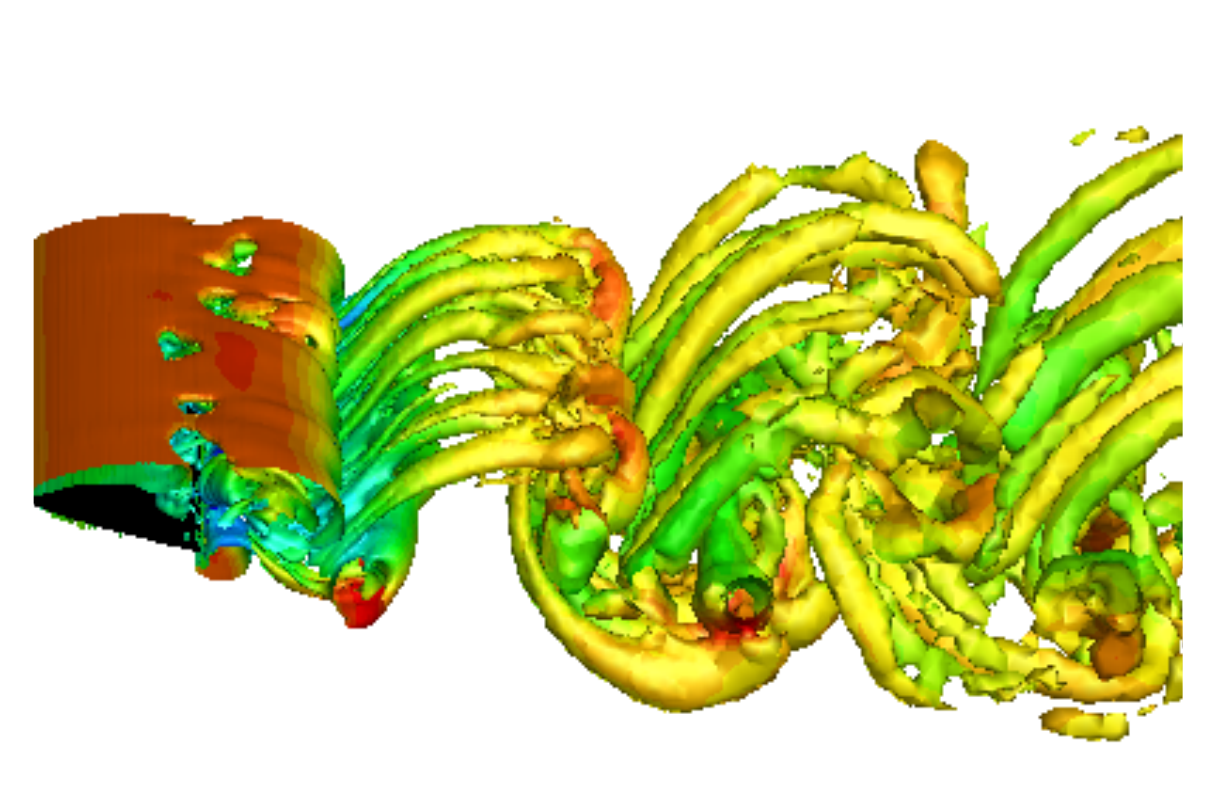} \\ \hline   
            	\begin{sideways}$\overline{u_x}$\end{sideways} & \includegraphics[width=0.31\textwidth]{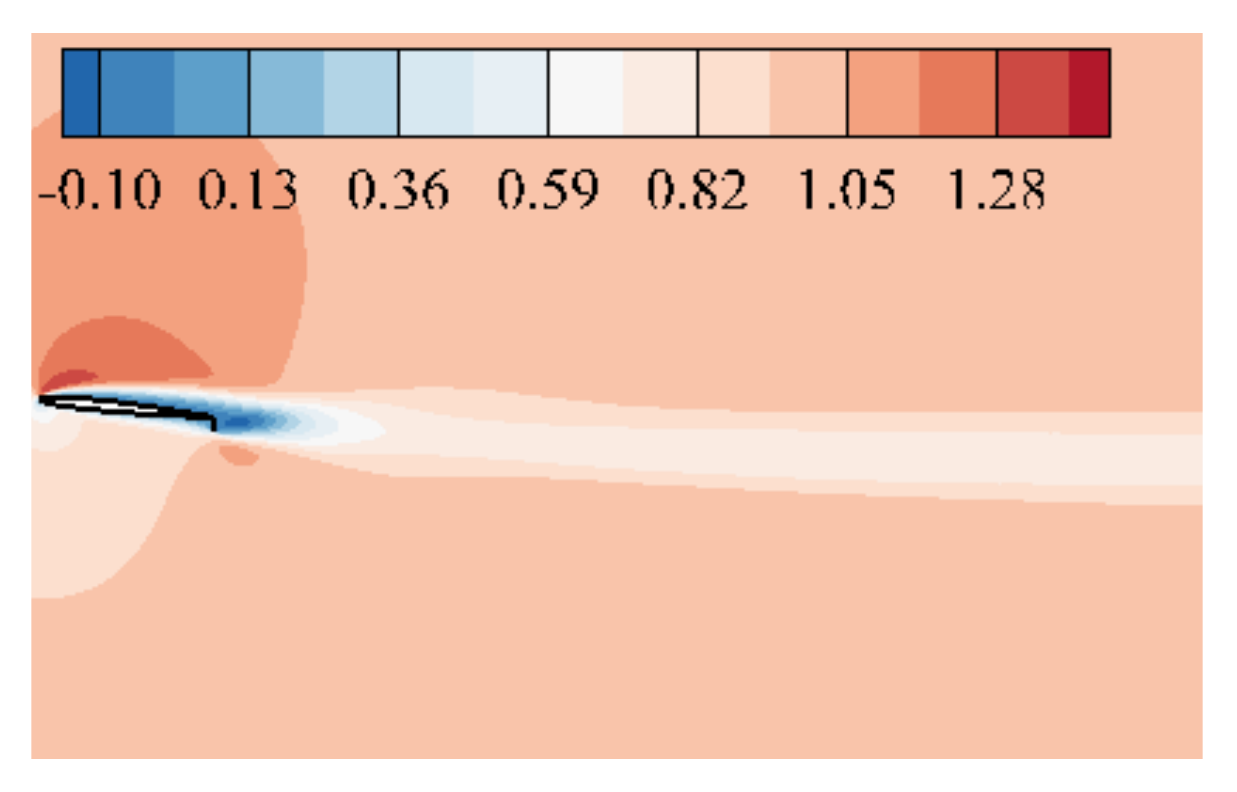} & \includegraphics[width=0.31\textwidth]{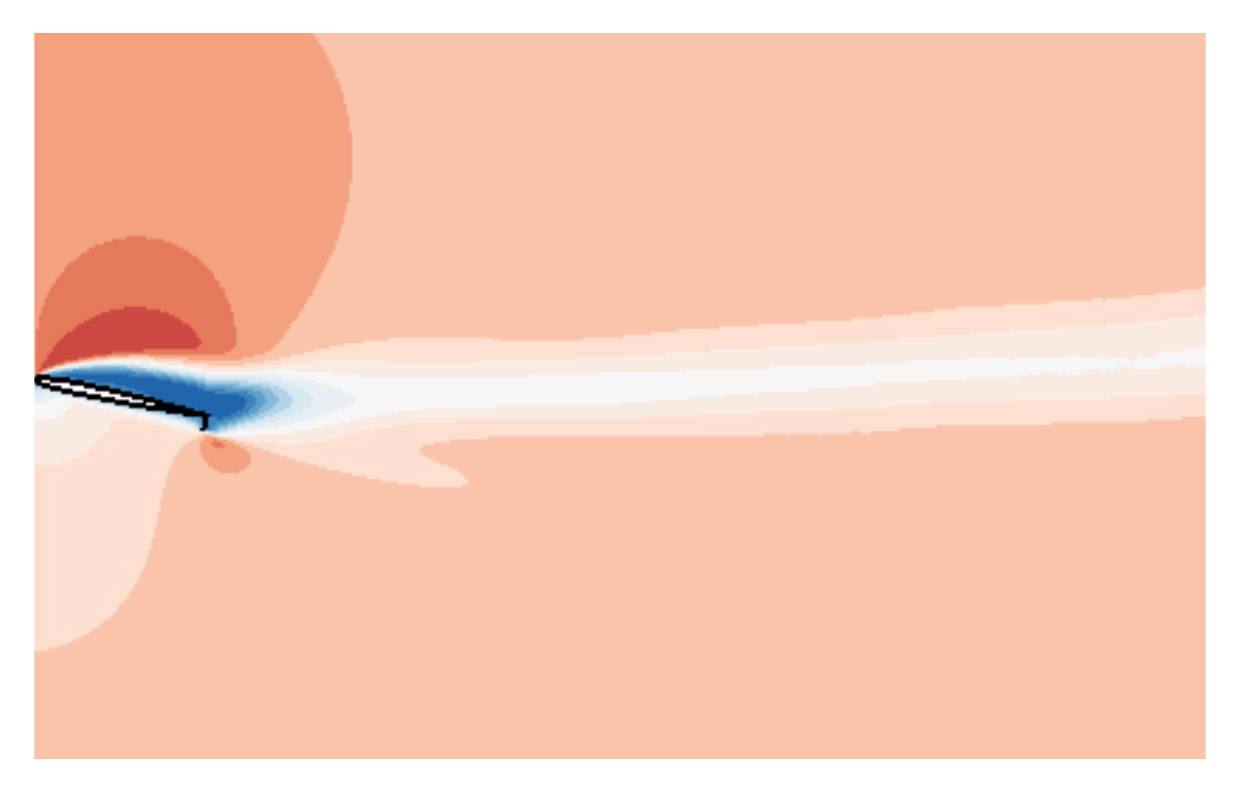} & \includegraphics[width=0.31\textwidth]{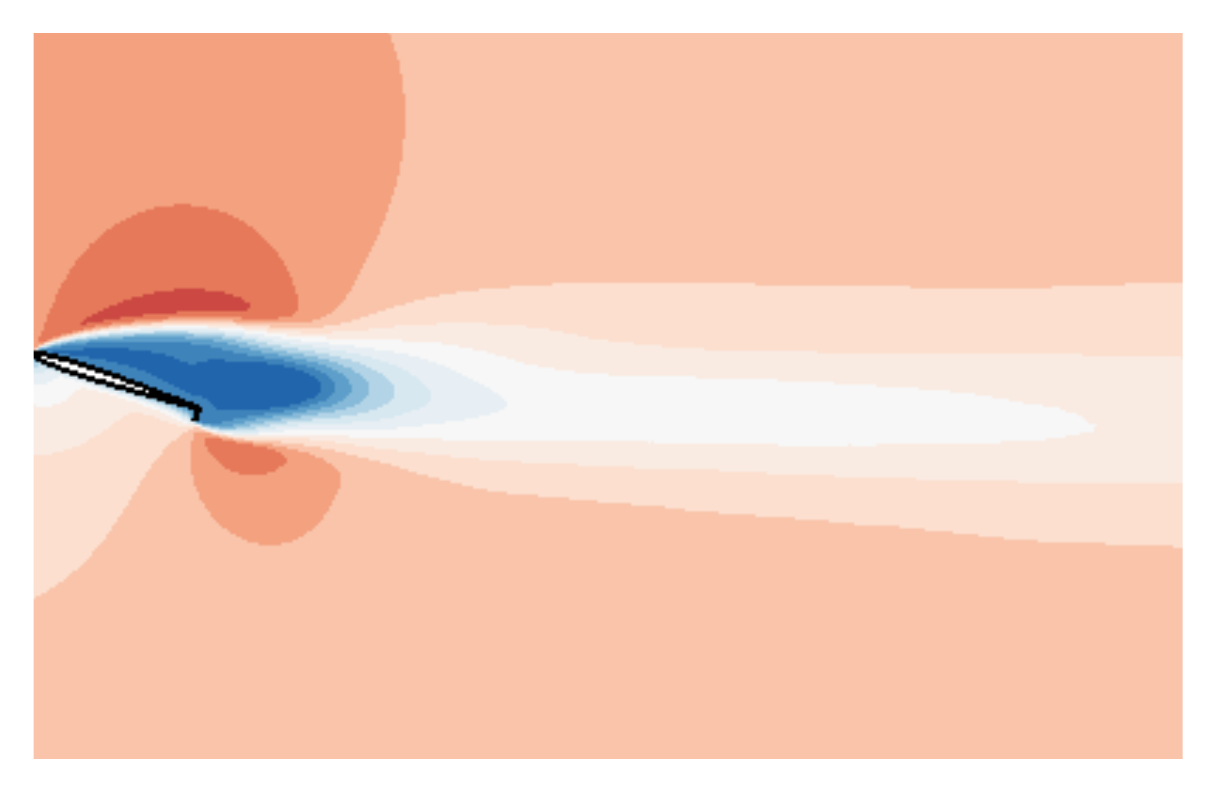} \\ \hline   
            	\begin{sideways}Force\end{sideways} & \includegraphics[width=0.22\textwidth]{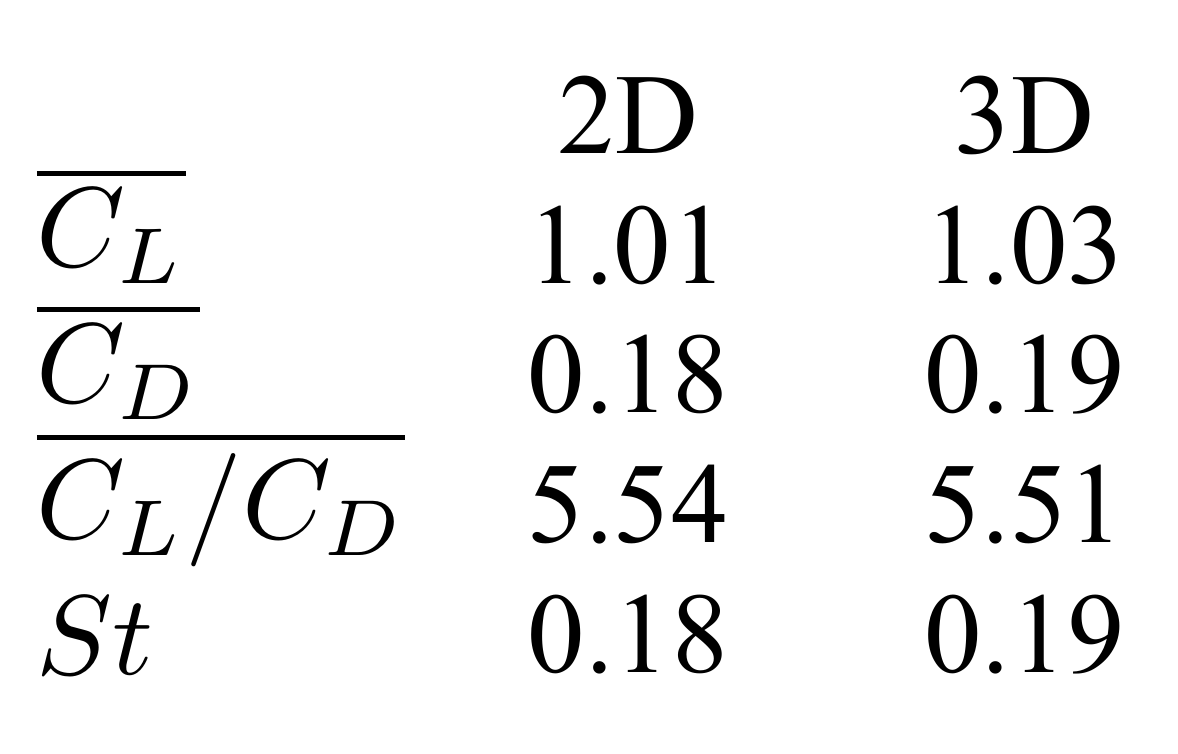} & \includegraphics[width=0.22\textwidth]{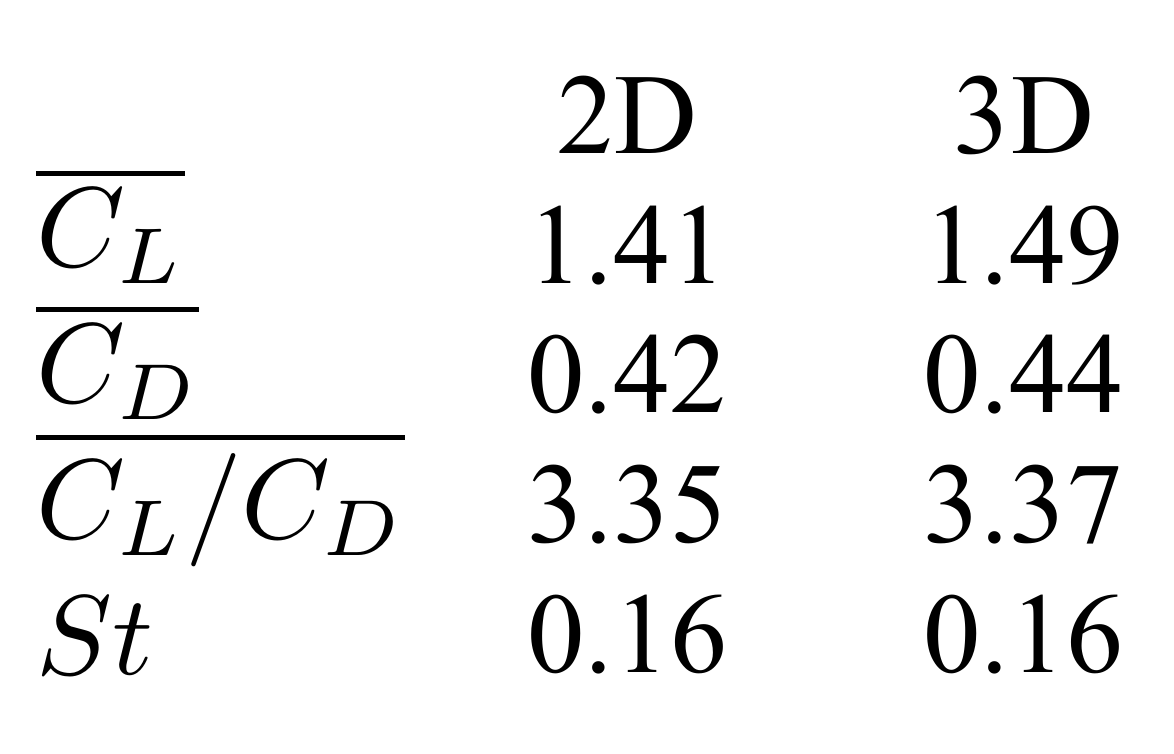} & \includegraphics[width=0.22\textwidth]{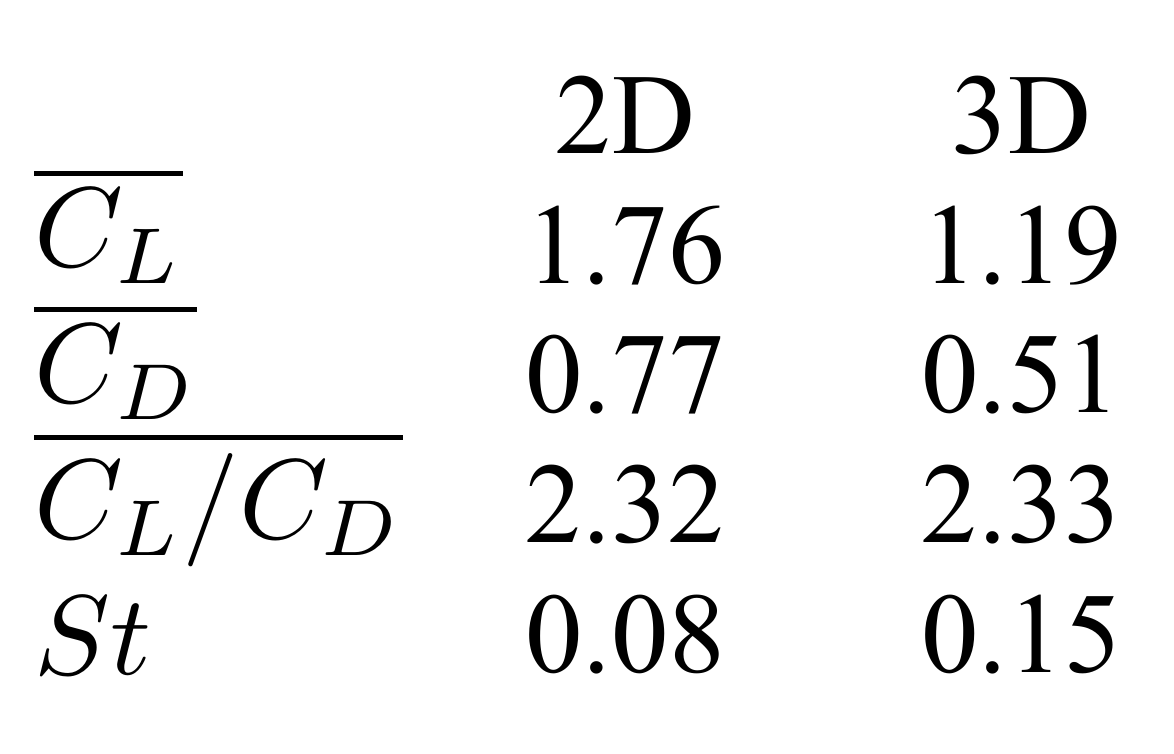} \\ \hline     
  \end{tabular}
\end{center}
\caption{Flow visualization and aerodynamic force characteristics of flow over the NACA 0006 airfoil with Gurney flap of $h/c = 0.08$ at $\alpha = 6^\circ, 12^\circ$ and $18^\circ$ with three-dimensional DNS. Each row consists of the top and perspective views of instantaneous snapshots of $Q$-criterion \cite{Hunt:CTR88} iso-surface colored with $u_x$, the time-averaged $\overline{u_x}$ and the comparison of aerodynamic force characteristics between the respective cases of two-dimensional and three-dimensional analysis.}
\label{t:3D_table}
\end{figure}

Three-dimensional analysis is performed on the flow around the NACA 0006 airfoil with Gurney flap of $h/c = 0.08$ at $\alpha = 6^\circ, 12^\circ$ and $18^\circ$ to examine the spanwise effects on the wakes. The chosen angles of attack yield the 2S, P and 2P wake regimes, respectively, for two-dimensional flow as shown in Fig.~\ref{f:stability}(b). The three-dimensional flow fields and the resulting force characteristics are summarized in Fig.~\ref{t:3D_table}. We utilize the iso-surface of $Q$-criterion \cite{Hunt:CTR88} to visualize the three-dimensional airfoil wake. The aerodynamic force characteristics of the three-dimensional flow are also compared to those obtained from the two-dimensional analysis.

The cases with $\alpha = 6^\circ$ and $12^\circ$, classified as the 2S and P wakes, respectively, in two-dimensional flow, remains two-dimensional with spanwise extension of the airfoil. Although the $12^\circ$ case shows some spanwise variations in the flow, it can be considered as nominally two-dimensional. The instantaneous snapshots of the two cases show the presence of the alternating positive and negative vortices leading to the formation of the K\'{a}rm\'{a}n vortex street, previously observed in the two-dimensional analysis. The time-averaged flow field also corresponds to that observed for a K\'{a}rm\'{a}n wake, and previously observed in the two-dimensional analysis. These observations suggest that the dominant vortex shedding frequency, lift and drag forces, and the lift-to-drag ratio experienced by the airfoil are similar to those observed in the two-dimensional analysis. This is noted from the similarity in the aerodynamic force characteristics for the two-dimensional and three-dimensional results in Fig.~\ref{t:3D_table}, further highlighting the two-dimensional nature of the flow. The same criteria used in the two-dimensional analysis are applied to classify the $\alpha = 6^\circ$ and $12^\circ$ cases under the 2S and P wake regimes.

The case with $\alpha = 18^\circ$, classified under the 2P regime for two-dimensional flow, is influenced by the spanwise instabilities in the flow and results in a fully three-dimensional flow with an underlying K\'{a}rm\'{a}n wake. The formation of the 2P wake mode is not observed. The 2P wake structure is disrupted by the spanwise instabilities and the wake structure is suppressed to a K\'{a}rm\'{a}n vortex street with spanwise variations. As seen from the top-view of Fig.~\ref{t:3D_table}, the cores of the positive and negative vortices in the K\'{a}rm\'{a}n street also show spanwise variations. The time-averaged flow field depicts the reduction in the wake thickness considerably compared to that observed in the two-dimensional 2P wake structure (see Fig.~\ref{t:regime}). This reflects in considerable reduction in the drag force experienced by the airfoil compared to that for the two-dimensional 2P regime. The altered dynamics of the shear layer roll-up leads to a reduction in the lift force compared to the two-dimensional 2P case. These inferences are observed from the aerodynamic force characteristics of the $\alpha = 18^\circ$ case in Fig.~\ref{t:3D_table}. The considerable decrease in the drag force compensates for the reduction in the lift enhancement experienced by the airfoil to maintain the lift-to-drag ratio. The dominant shedding frequency of $St = 0.155$, similar to the values observed for that of the 2S and P wake structures, substantiates the K\'{a}rm\'{a}n wake behavior for the dominant vortical structures of the three-dimensional flow field.

We have observed that the two-dimensional analysis provides a foundation to understand the occurrence of the 2S and P wake regimes for the three-dimensional spanwise-periodic setup. The deviation with three-dimensional flow appears for the fully three-dimensional flow for what would be the 2P wake regime in the two-dimensional setting. Nonetheless, it should be noted that at higher $Re$, three-dimensionality of the wake structures can be instigated to even the flow classified as two-dimensional wake regimes at this $Re$. This transition of the flow field of each particular wake regime from two-dimensional to three-dimensional is beyond the scope of the current work, and the readers are referred to studies which focus in such analyses \cite{Freymuth:66,Braza:JFM01,Hoarau:JFM03}. Further analyses have to be performed to completely understand these effects for the current problem. The novelty of the current work is in understanding the transition of the wake regimes from steady through 2P or K\'{a}rm\'{a}n vortex street having spanwise variations with change in different parameters of airfoil thickness, angle of attack, effects of the Gurney flap and the flap height, and associated effects on the force characteristics of the airfoil in an incompressible flow at $Re = 1000$.

\section{Concluding Remarks}
We perform a large number of two-dimensional direct numerical simulations for incompressible flow over different airfoils - NACA 0000 (flat plate), 0006, 0012 and 0018, at $Re = 1000$ with and without a Gurney flap attached to the trailing edge over a range of angles of attack. Use of the Gurney flap significantly enhances the lift experienced by the airfoils. Lift-to-drag ratio increase exceeding twice that of the baseline cases are observed with the attachment of the Gurney flap. However, the benefit of lift enhancement is accompanied with a penalty of drag increase, particularly at high angles of attack. This further leads to degradation of the lift-to-drag ratios below baseline cases at high angles of attack, although high lift enhancement is also observed at these flow conditions. The overshadowing effect of thicker airfoils on the Gurney flap reduces the influence of the flap on flow modification in such airfoils. The aerodynamic forces experienced by these airfoils are also significantly altered compared to that for thinner airfoils. Even though the magnitude of lift and lift-to-drag ratio enhancement is reduced for thicker airfoils, the lift as well as the drag slopes are much smoother and the drag is also reduced at higher angles of attack. 

Analyzing the lift spectra and flow structure of the airfoil wake reveals the presence of different vortex shedding patterns in the wake. These are classified into four wake regimes: (1) steady, (2) 2S - periodic von K\'{a}rm\'{a}n vortex shedding of alternating counter-clockwise and clockwise rotating vortices, (3) P - periodic von K\'{a}rm\'{a}n shedding of a single vortex pair, and (4) 2P - periodic shedding of two distinct vortex pairs. Characteristic Strouhal number corresponding to the vortex shedding phenomena is also observed for each of the wake modes. Detailed analysis of the near wake properties also revealed the wake mode transitions. Among the different vortical structures formed in the near wake of the airfoils, the observations suggest that the strong LEV and the secondary vortical structure with counter-clockwise direction of vorticity formed on the suction side of the airfoil lead to the roll up of the trailing-edge vortex onto the suction side at high angles of attack and Gurney flap heights. This leads to the emergence of different vortex shedding patterns and, hence, the wake mode transitions. We also perform a selected number of three-dimensional DNS of representative cases to analyze the three-dimensional effects on these wake mode transitions in a spanwise periodic setting. The results suggest that the 2S and P wake modes tend to remain two-dimensional even with spanwise effects, although the flow presents three-dimensionality at very high angles of attack at this $Re$ of 1000. For a three-dimensional flow, the 2P mode is suppressed by the spanwise instabilities reducing to a K\'{a}rm\'{a}n vortex street like wake structure, suggesting the need for careful three-dimensional analysis.

The existence of the wake modes is summarized in a wake classification diagram over a wide range of angles of attack and Gurney flap heights for each of the airfoils. Unsteady wake regimes appear at lower angles of attack with the addition of the Gurney flap. The wake classification diagram provides guidance on the implementation of the Gurney flap at different conditions for the required aerodynamic performance, motivating studies in passive as well as active Gurney flaps. The airfoil performance can be assessed depending on the nature of wake regime the respective flow is classified under for the particular angle of attack and flap height.


\section*{Acknowledgments}
The authors thank the insightful discussions with Dr.~Phillip Munday, Chi-An Yeh and Daisuke Oshiyama. The majority of the computation for this project was performed at the Research Computing Center at the Florida State University. MGM and KT were partially supported by the National Science Foundation (Award Number 1632003).


\bibliography{refs}   
\bibliographystyle{aiaa}

\end{document}